\documentclass{jfm}
\usepackage{multirow}
\usepackage{graphicx}
\usepackage{amsmath}
\usepackage{epstopdf}
\usepackage{amssymb}
\usepackage{dcolumn}
\usepackage{bm}
\usepackage{booktabs}
\usepackage{overpic}
\usepackage[colorinlistoftodos,prependcaption,textsize=tiny,shadow]{todonotes}

\graphicspath{{Figures/}}

\shorttitle{Hypersonic attachment-line instabilities}
\shortauthor{Y. C. Xi, J. Ren and S. Fu}

\title{Hypersonic attachment-line instabilities with large sweep Mach numbers}

\author{Youcheng Xi\aff{1}, Jie Ren\aff{2}
 \and Song Fu\aff{1}\corresp{\email{fs-dem@tsinghua.edu.cn}}}

\affiliation{\aff{1}School of Aerospace Engineering, Tsinghua University, Beijing 100084, China
\aff{2}Department of Mechanical Engineering, Faculty of Engineering, University of Nottingham, Nottingham, NG7 2RD, UK}

\begin{document}

\maketitle

\begin{abstract}
This study aims to shed light on hypersonic attachment-line instabilities with large sweep Mach numbers. Highly swept flows over a cold cylinder that give rise to large sweep Mach numbers are studied. High fidelity base flows are obtained by solving full Navier-Stokes equations with a high-order shock-fitting method. Using local and global stability theories, an attachment-line mode is found to be dominant for the laminar-turbulent transition along the leading edge that agrees well with experimental observations \citep{Gaillard1999}. The behavior of this mode explains the reason why the transition occurs earlier as the sweep Mach number is above 5.  Also, this attachment-line mode is absent if the base flow is calculated with boundary-layer assumptions, indicating that the influence of inviscid flow outside the boundary layer can not be ignored as is normally done.
It is clearly demonstrated that the global modes display the features of both attachment-line modes, as in sweep Hiemenz flow, and cross-flow-like modes further downstream along the surface. In contrast to incompressible flows, 
the mode is shown to be of inviscid nature. Moreover, the leading-edge curvature has a destabilizing effect on the attachment-line mode for large spanwise wave numbers but a stabilizing effect for small spanwise wave numbers. 
\end{abstract}


\section{Introduction}\label{S1}

The mechanism of boundary layer transition is one of the most active research fields in contemporary fluid dynamics. Not only because of its complexity in mathematics and physics but also for the enormous potential applications in practical engineering.  Over the years, linear stability theory (LST) plays an essential role in revealing mechanisms of flow instability. Moreover, by using LST, some of the fundamental mechanisms are now well understood. Representative examples are Tollmien-Schlichting (TS) waves \citep{Schlichting2017}, Mack modes \citep{Mack1975} and cross-flow modes \citep{Saric2003} in two and/or three-dimensional boundary layers. More detailed works in this field could be found in reviews by \citet{Reed1989,Reed1996}, \citet{Fedorov2011} and \citet{Zhong2012}. However, due to the richness of flow physics in high-speed flows, flow instability is still far from fully understood, even in terms of fundamental modal instability.

In particular, the leading edge of a wing plays a very important role in boundary layer transition. One noticeable phenomenon in experiments is the leading edge contamination: if the Reynolds number is sufficiently high, initial turbulent flow could persist along the attachment line. In real flight vehicles, due to significant geometric variation in wing-body junctions, initial laminar flow could easily become turbulent,  contaminating the flow state of the attachment line. Such phenomenon motivated people to understand the mechanism of the attachment-line transition.

The steady laminar flow in the leading-edge region of a swept wing was often studied using Hiemenz model \citep{Rosenhead1963}. Unlike the similarity solution for conventional boundary layer flows, Hiemenz flow is an exact solution of incompressible Navier-Stokes equations. In fact, good agreement was achieved with this model as compared with experiments \citep{Gaster1967}. Thus, experimental and theoretical studies based on the Hiemenz model became popular. The works by \citet{Pfenninger1965}, \citet{Poll1979}, \citet{Lin1996,Lin1997}, \citet{Theofilis1995,Theofilis1998,Theofilis2003J} and \citet{Obrist2003a,Obrist2003b} were probably the most representative. Further, the stability feature of the attachment-line flow had also been investigated through direct numerical simulations (DNS) \citep{Spalart1988}. The numerical results confirmed that the leading unstable mode satisfied the assumption made by G{\"o}rtler \citep{Gortler1955dreidimensionale} and H{\"a}mmerlin \citep{Hammerlin1955instabilitatstheorie} for Hiemenz flow. Under this assumption, the linear instability in the attachment-line acquires the symmetry of the base flow, in which chord-wise velocity is a linear function of the chord-wise coordinate. \citet{Joslin1995} also performed DNS to study the behavior of perturbations along the attachment line and found the stabilizing effect of surface suction.

Early stability properties of subsonic compressible leading-edge boundary layer flow were discussed by \citet{Theofilis2006}. In their work, the problem was solved both numerically and theoretically. They demonstrated that the three-dimensional polynomial eigenmodes of an 
incompressible flow \citep{Theofilis2003J} persisted in the subsonic flow regime. Later, a more accurate analogy analysis based on sparse techniques was performed by \citet{Gennaro2013}. Their results perfectly matched those from theoretical analysis over a large parameter range in the subsonic region. They found that when the sweep Mach number decreased, the range of unstable region and the growth rate became larger, but the critical Reynolds numbers increased.

As the free-stream Mach number further increases from subsonic to supersonic,  the compressibility effects become more significant.
The investigation of supersonic attachment-line flow was initially focused on the influence of sweep angle and the heat flux along the attachment line \citep{Gallagher1959}. The transition of the attachment line flow was also detected by \citet{Gallagher1959}. In their Mach 4.15 experimental study, the effect of sweep angles was studied in a relatively large range.  Later, \citet{Creel1986} performed experiments with free-stream Mach number of 3.5 and several sweep angles. They also detected transition along the attachment-line and the critical transition Reynolds numbers to be around 650. \citet{Skuratov1991} performed similar test to validate Creel {\it et al.}'s result. \citet{Murakami1996} conducted experiments on hypersonic attachment-line flow in Ludwieg-tube wind tunnel. They found that the critical Reynolds number increased slightly as the sweep Mach number increased. \citet{Gaillard1999} presented extensive experimental results for hypersonic attachment-line flow with various sweep Mach numbers. It is interesting to note that the critical Reynolds number decreased as the sweep Mach number was above 5.

Apart from experimental studies, researchers also tried to understand the Mach number effect theoretically.
An early theoretical attempt to study the stability of compressible attachment line was made by \citet{Malik1988} with perturbations of T-S type. But this assumption neglected the chord-wise dependence of the base flow. A more proper assumption was made later by \citet{Lin1995}, where two-dimensional eigenvalue problems were directly solved, allowing two-dimensional dependence of the mean flow in the solution. It was found that the attachment-line flow was subject to three-dimensional instability. Also, the critical Reynolds number based on the momentum thickness was found to be around 125. \citet{Semisynov2003} performed a combined theoretical and experimental study and found the critical transition Reynolds numbers were higher in supersonic than in subsonic flows. More recently, Mack {\it et al.} published a series of studies \citep{Mack2008, Mack2010a, Mack2011a, Mack2011b} focusing on hypersonic flows around a yawed parabolic body of infinite span with their innovatively developed Jacobi-free global stability solver \citep{Mack2010}. The global spectrum that contained both attachment-line and cross-flow instabilities was presented for sweep Mach number of 1.25. Some modes were found to reflect features of both attachment-line and crossflow instabilities. They also observed that the unstable acoustic mode could coexist with the unstable boundary mode. The relative critical Reynolds number of the most unstable acoustic mode was smaller than the unstable boundary mode. 


From the above reviews, the attachment-line instability is still not clearly understood, most prominently in the hypersonic region where sweep Mach number is considerable, as highlighted in the series of experiments from \citet{Gaillard1999}. In this region, no theoretical explanation is present to explain why the critical Reynolds number decreased when the sweep Mach number was above 5. Also, the curvature effect, the nature of unstable modes  are not studied under this condition. This study provides a comprehensive analysis using both local and global stability theory in an attempt to uncover the transition mechanisms related to large sweep Mach numbers.


In Section~\ref{S2} the methodologies for base flow and stability analysis are introduced. The base flow is discussed in Section~\ref{S3}. In Section~\ref{S4} the local analysis and global analysis are discussed. The paper is concluded in Section~\ref{S5}.

\section{Methodology and Problem Formulation}\label{S2}
\subsection{Description of the Problem}\label{S2.1}
\begin{figure}
\begin{center}
\includegraphics[width=0.65\textwidth]{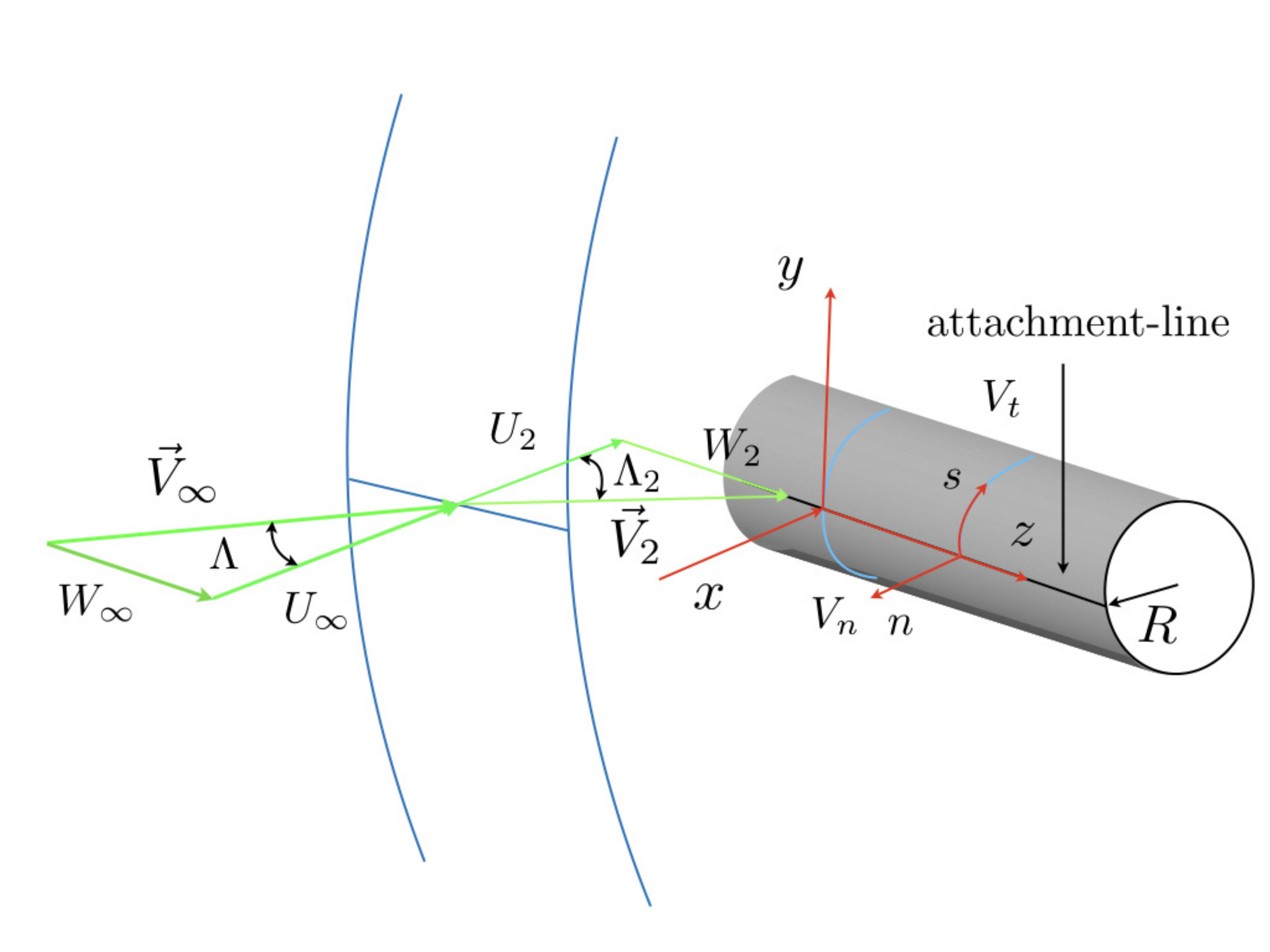}
\caption{Schematic of hypersonic flow around an inclined cylinder. The velocity vector ahead of the shock is \(\vec{\bm{V}}_{\infty}=(U_{\infty},V_{\infty},W_{\infty})\) and \(\vec{\bm{V}}_2 = (U_{2},V_{2},W_{2})\) is the velocity vector behind the shock. \(V_t\) and \(V_n\) represent the velocity along the surface tangential direction \(s\) and wall normal direction \(n\). \(\Lambda\) and \(\Lambda_2\) represent the sweep angles in the freestream and behind the shock.}
\label{Fig1}
\end{center}
\end{figure}

The hypersonic flow around a swept cylinder is studied here based on relevant experimental conditions \citep{Gaillard1999}. As shown in figure \ref{Fig1}, a cylinder of radius \(R=33mm\) is assumed to be of infinite length in the spanwise direction. The incoming flow impinges onto the surface of the cylinder with a sweep angle \(\Lambda\). Flow parameters before and after the shock wave, denoted with subscripts \(\infty\) and \(2\) respectively,  satisfy the Rankine-Hugoniot (R-H) relations. Velocity components \(U\),\(V\) and \(W\), are defined along the \(x\),\(y\) and spanwise \(z\) axis of the Cartesian coordinates.  The subscripts \(s\) and \(n\) are used to represent the surface and wall normal directions, along which the velocities are denoted with \(V_t\) and \(V_n\), respectively.

Following the previous studies by \citet{Mack2008,Mack2011a}, we define  a free-stream Reynolds number \(Re_{\infty}\), a sweep Reynolds number \(Re_s\), a free-stream Mach number \(M_{\infty}\), a sweep Mach number \(M_s\) and a recovery temperature \(T_r\) as,

\begin{equation} \label{eq1}
\begin{aligned}
Re_{\infty} = \frac{|\vec{V}_{\infty}| R}{\nu_{\infty}},~Re_s = \frac{W_2 \delta}{\nu_r},~M_{\infty} = \frac{|\vec{V}_{\infty}| }{c_{\infty}}, ~M_s = \frac{W_2}{c_2}, \\
 T_r = T_{\infty} + \sigma (T_0 - T_{\infty}), \textrm{where}~\sigma = 1 - (1 - \xi_w) sin^2\Lambda.
 \end{aligned}
\end{equation}
In \eqref{eq1}, $\xi_w$ is a constant for specific free stream conditions (\(M_{\infty}\) and \(\Lambda\)) 
 and determined based on the study of \citet{Reshotko1958}. The parameter $c$ is the speed of sound, \(\nu_r\) represents the kinematic viscosity at the recovery temperature $T_r$. The viscosity lengths scale $\delta$ is determined as
\begin{equation}
\delta =\sqrt{\frac{\nu_r R}{2U_2}}. 
\label{eq2}
\end{equation}
The free-stream Mach number $M_\infty=7.14$ and temperature \(T_{\infty} = 69.84K\) are fixed for all cases. A cold wall temperature is specified as $T_w = 0.4 T_r$ according to experimental conditions. The Prandtl number \(Pr=0.72\) and the specific heat ratio \(\gamma=1.4\) is defined following the ideal gas assumption of air. The defined six cases are shown in table \ref{table1}. For each case, the viscosity length scale $\delta$ is different due to the disparity in sweep angle. 
Since most high-order finite difference methods \citep{Mack2010a,Zhong1998,Lele1992} for solving NS/Euler equations need to reduce the scheme order at boundary regions to satisfy the dissipation-error and stability conditions, the full-size model is used to maintain the scheme order around attachment-line, even though the flow is symmetric to the \(x-z\) plane at \(y=0\) at zero angle of attack. Many methods can be used to obtain the laminar base flow. The most appropriate one is the Direct-numerical-simulation(DNS) approach by solving the Navier-Stokes (NS) equation with high-order shock fitting methods \citep{Moretti1987,Kopriva1999,Zhong1998} which take all the information into account. The other is the combination of solving inviscid Euler equation and boundary layer equations (see \citet{Wang2018} and \citet{Theofilis2006} for more details about solving boundary layer equations), which is much cheaper but overlooks the influence of the inviscid flow outside the boundary layer. Both methods (DNS and boundary layer assumption) are used and compared in the present study.

\begin{table}
\begin{center}
\setlength{\tabcolsep}{0.007\textwidth}
{
\begin{tabular}{c|ccccccccccc}
Case &  $M_{\infty}$ & $\Lambda(^o)$ & $T_{\infty}(K)$ & $\delta(m)$ &  $T_r/T_{\infty}$ & $T_w/T_{\infty}$ & $\rho_r/\rho_{\infty}$ & $Re_{\infty} $ & $Re_{s} $ & $M_s$ & $R/\delta$\\ 
\hline
C3376a & 7.14 & 76.5  & 69.84 & 1.4937e-4 &   9.89  & 3.95  & 2.62     & 2704.73   &  986.04  &5.8      &  220.93\\ 
C3375   & 7.14 & 75     & 69.84 & 1.4400e-4 &   9.89  & 3.95  & 2.91     & 2601.36   & 1043.68 &5.51      &  229.17\\
C3374 & 7.14 & 74     & 69.84 & 1.4000e-4 &   9.90  & 3.95  & 3.09      & 2537.43  & 1075.44 &5.32     &  235.71 \\
C3373 & 7.14 &  73    & 69.84 & 1.3600e-4 &   9.90  & 3.96  & 3.26      & 2477.43  & 1102.14 &5.15      & 242.65 \\
C3370   & 7.14 & 70     & 69.84 & 1.2800e-4 &   9.94  & 3.98  & 3.72      & 2319.10  & 1155.19 &4.65      & 257.81 \\
C3365   & 7.14 &  65    & 69.84 & 1.1700e-4 &  10.01 & 4.00  & 4.32      & 2113.39  & 1174.21 &3.94      & 282.05 \\
\end{tabular}}
\end{center}
\caption{Parameters of the flow cases in the current study. The names of cases are the same as in experiment\citep{Gaillard1999}. The 'C' represents the Cylinder. The first two number represent the radius and the last two numbers represent the sweep angle. \(\rho_r\) represents the density of the fluid at the recovery temperature \(T_r\) and \(\rho_{\infty}\) represents the density of free stream.}
\label{table1}
\end{table}%

\subsection{Mathematical Formaultion}
\subsubsection{Flow governing equations}
The problem solution starts from the unsteady three-dimensional N-S equations:
\begin{subequations}
\begin{equation}
\frac{\partial Q}{\partial t}+\frac{\partial F_j}{\partial x_j} + \frac{\partial F_{vj}}{\partial x_j}=0,
\end{equation}
\begin{equation}
Q = \left[ {\begin{array}{*{20}{c}}
  \rho  \\ 
  {\rho {u_1}} \\ 
  {\rho {u_2}} \\ 
  {\rho {u_3}} \\ 
  {{E_t}} 
\end{array}} \right],{F_j} = \left[ {\begin{array}{*{20}{c}}
  {\rho {u_j}} \\ 
  {\rho {u_1}{u_j} + p{\delta _{1j}}} \\ 
  {\rho {u_2}{u_j} + p{\delta _{2j}}} \\ 
  {\rho {u_3}{u_j} + p{\delta _{3j}}} \\ 
  {\left( {{E_t} + p} \right){u_j}} 
\end{array}} \right],{F_{vj}} = \left[ {\begin{array}{*{20}{c}}
  0 \\ 
  {{\tau _{1j}}} \\ 
  {{\tau _{2j}}} \\ 
  {{\tau _{3j}}} \\ 
  {{\tau _{jk}}{u_k} - {q_j}} 
\end{array}} \right],
\end{equation}
\end{subequations}
The total energy \(E_t\) and the viscous stress \(\tau_{ij}\) are given as, respectively, 
\begin{equation}
E_{t}=\rho\left(\frac{T}{\gamma(\gamma-1) M^{2}}+\frac{u_{k} u_{k}}{2}\right), \quad \tau_{i j}=\frac{\mu}{Re_{\infty}}\left(\frac{\partial u_{i}}{\partial x_{j}}+\frac{\partial u_{j}}{\partial x_{i}}-\frac{2}{3} \delta_{i j} \frac{\partial u_{k}}{\partial x_{k}}\right).
\end{equation}
The pressure \(p\) and heat flux \(q_i\) are obtained from:
\begin{equation}
p=\frac{\rho T}{\gamma M_{\infty}^{2}}, \quad q_{i}=-\frac{\mu}{(\gamma-1) M_{\infty}^{2} Re Pr} \frac{\partial T}{\partial x_{i}}.
\end{equation}
The viscosity is calculated using the Sutherland law
\begin{equation}
\mu = T^{3/2}\frac{T_{\infty} + C}{T T_{\infty} + C},
\end{equation}
with \(C = 110.4K\). The fifth-order upwind scheme (for inviscid flux \(F_{j}\)) of \citet{Zhong1998} together with the six-order center scheme (for viscous flux \(F_{vj}\)) is used to compute the flow field. Here, the non-conservative characteristic relation is adapted at the shock surface for more convenient stability analysis. A 4th-order Runge-Kutta method is used to perform the time integration. By treating the shock wave as a sharp interface, high accuracy can be achieved in the whole flow field, which is an essential prerequisite for the stability analysis. The Euler equation is solved by the same method by ignoring the viscous flux \(F_{vj}\).

Once the base flow is obtained, the linear Navier-Stokes (LNS) equation of the perturbations are solved. The LNS equations are derived from the NS equations by introducing small perturbations, subtracting the base flow equations and ignoring the nonlinear terms. A frequently-employed form is commonly written as 
\begin{align}   \label{eq7}
\bm{\Gamma}\frac{\partial \bold{\Phi}}{\partial t} +\mathbf{A}\frac{\partial \bold{\Phi}}{\partial x}
&+\mathbf{B}\frac{\partial \bold{\Phi}}{\partial y} +\mathbf{C}\frac{\partial \bold{\Phi}}{\partial z}+\mathbf{D}\bold{\Phi}=  \\ \nonumber
&\mathbf{H}_{xx}\frac{\partial^2 \bold{\Phi}}{\partial x^2}+\mathbf{H}_{xy}\frac{\partial^2 \bold{\Phi}}{\partial x\partial y}
+\mathbf{H}_{xz}\frac{\partial^2 \bold{\Phi}}{\partial x\partial z}+\mathbf{H}_{yy}\frac{\partial^2 \bold{\Phi}}{\partial y^2}
+\mathbf{H}_{yz}\frac{\partial^2 \bold{\Phi}}{\partial y\partial z}+\mathbf{H}_{zz}\frac{\partial^2 \bold{\Phi}}{\partial z^2},
\end{align}
where the coefficient matrix \(\bm{\Gamma},\mathbf{A},\mathbf{B},\mathbf{C},\mathbf{D},\mathbf{H}_{xx},\mathbf{H}_{xy},\mathbf{H}_{xz},\mathbf{H}_{yy},\mathbf{H}_{yz},\mathbf{H}_{zz}\) can be found in \citep{Ren2014,Ren2015,Wang2017}.

\subsubsection{Linear local stability approach}
For local analysis, the perturbations along the attachment-line can be written in a wavelike form as:
\begin{equation} \label{eq8}
\bold{\Phi}(x,y,z,t) = \vec{\phi}(x)\text{exp}\left(i\alpha y + i\beta z - i\omega t\right) + c.c., 
\end{equation}
where \(\vec{\phi} = (\rho',u',v',w',T')\) is the shape function, \(\alpha\) and \(\beta\) are the wavenumbers along \(y\) and \(z\) directions, \(\omega\) is the frequency and \(c.c.\) represents the complex conjugate. Since \(\alpha\) is not known a priori for local calculations, a two-dimensional perturbation wave assumption is used here as \(\alpha = 0\). Substituting \eqref{eq8} into \eqref{eq7}, the LNS reduces to a generalized eigenvalue problem as

\begin{equation}
\mathfrak{L}_l\vec{\phi} = \omega\mathfrak{R}_l\vec{\phi},
\end{equation}
where \(\mathfrak{L}_l\) and \(\mathfrak{R}_l\) are matrix operators: 
\begin{align}
\mathfrak{L}_l=\left(\mathbf{D} + i\beta \mathbf{C} + \beta^2\mathbf{H}_{zz}\right) +\left(\mathbf{A}-i\beta \mathbf{H}_{xz}\right)\frac{\partial}{\partial x} - \mathbf{H}_{xx}\frac{\partial^2}{\partial x^2},
\end{align}
\begin{align}
\mathfrak{R}_l =i\bm{\Gamma}.
\end{align}
A temporal stability analysis is performed considering the homogeneous nature in the spanwise direction. In the wall normal direction, grids cluster near the wall surface in the following manner
\begin{equation} \label{eq9}
y=a\frac{1+\eta}{b-\eta},\quad\text{with}\quad a=\frac{y_iy_{max}}{y_{max}-2y_{i}},b=1+\frac{2a}{y_{max}},\eta\in\left[-1,1\right],
\end{equation}
where \(y_{max}\) represents the far-field and \(y_i\) is the control point.
This grid distribution allows for clustering of half of the grid points in the region \(\left[0,y_i\right]\), similar to the work of \citet{Schmid2001}.  The spectral method is used for approximation of the derivatives and a standard QZ solver\citep{Golub2013} is used for solving the eigenvalue problems.

\subsubsection{Global stability approach}
From the global point of view, perturbations can be written in a more general form:
\begin{equation} \label{eq10}
\bold{\Phi}(x,y,z,t) = \vec{\phi}(x,y)\exp\left(i\beta z - i\omega t\right) + c.c. 
\end{equation}
Substituting \eqref{eq10} into \eqref{eq7}, one again arrives at a generalized eigenvalue problem,
\begin{equation} \label{eq11}
\mathfrak{L}\vec{\phi} = \omega\mathfrak{R}\vec{\phi},
\end{equation}
where \(\mathfrak{L}\) and \(\mathfrak{R}\) are matrix operators: 
\begin{align}
\mathfrak{L}=\left(\mathbf{D}+i\beta \mathbf{C} + \beta^2\mathbf{H}_{zz}\right)&+
\left(\mathbf{A}-i\beta \mathbf{H}_{xz}\right)\frac{\partial}{\partial x}+
\left(\mathbf{B}-i\beta \mathbf{H}_{yz}\right)\frac{\partial}{\partial y}  \\ \nonumber
&-\mathbf{H}_{yy}\frac{\partial^2}{\partial y^2}-\mathbf{H}_{xy}\frac{\partial^2}{\partial x\partial y} - \mathbf{H}_{xx}\frac{\partial^2}{\partial x^2},
\end{align}
\begin{equation}
\mathfrak{R}=i\bm{\Gamma}.
\end{equation}
Considering the length scale of instability of the previous study \citep{Mack2008, Mack2010a, Mack2011a, Mack2011b}, the basic nondimensional spanwise wave number \(\beta\) is chosen to be \(0.3\). 
Because the matrices discretizing the global stability problem have leading dimensions of \(O(10^5-10^6)\), instead of classical QZ method, a Krylov-Shur method \citep{Stewart2002b,Stewart2002a}, based on PETSc ({http://www.mcs.anl.gov/petsc}) and SLEPc ({http://slepc.upv.es}) with various spectral transformation techniques have been used to recover a window (100 - 400) of the eigenvalues of interest. The Krylov-Shur method, which is another kind of implicitly restarted Arnoldi algorithm, can achieve very high precision for specific part of the spectrum with proper spectral transformations. Sparse linear algebra packages, MUMPS ({http://mumps.enseeiht.fr}) and SuperLU ({http://crd-legacy.lbl.gov/~xiaoye/SuperLU/}) are used to undertake the inverse of a matrix during the spectral transformations.
 In both directions, special mesh distribution (FD-q grids) based on \citet{Hermanns2008} is implemented according to the order of the scheme as first discussed by \citet{Paredes2013}. Again, FD-q grids cluster near the wall surface using equation \eqref{eq9} and an 8-th order FD-q scheme is used.

\subsubsection{Boundary conditions}
In the base flow, no slip boundary condition together with the isothermal wall on the cylinder surface are employed. At the end of chord-wise or surface tangential direction for the computational domain, characteristic non-reflect boundary conditions are imposed. In the calculation of perturbations, no slip and Dirichlet conditions for temperature are specified at the wall (\((u',v',w',T')= 0\)). At the far field, along the shock surface, all perturbations except density are forced to zeros. Along the \(s\) direction, at the exit, a high-order extrapolation is performed from interior for all perturbation quantities.
 
\section{Base flow}\label{S3}

The present analysis covers sweep Mach number roughly from 4 to 6 as shown in subsection \ref{S2.1}.
Due to the discrepancy in shock shapes, computation domains are therefore different among cases as listed in table \ref{table1}. For all cases, a mesh is generated with: 641 grids points in the surface direction (clustered around the leading edge), 221 grids points in the wall-normal direction (at least 35 points clustered inside the boundary layer) and 8 grid points in the spanwise direction due to the homogeneous nature in this flow. Compared with previous DNS  study \citep{Mack2008,Speer2004}, the base flow can be adequately resolved under this grid resolution. 

\begin{figure}
\begin{center}
 \includegraphics[width=0.9\textwidth]{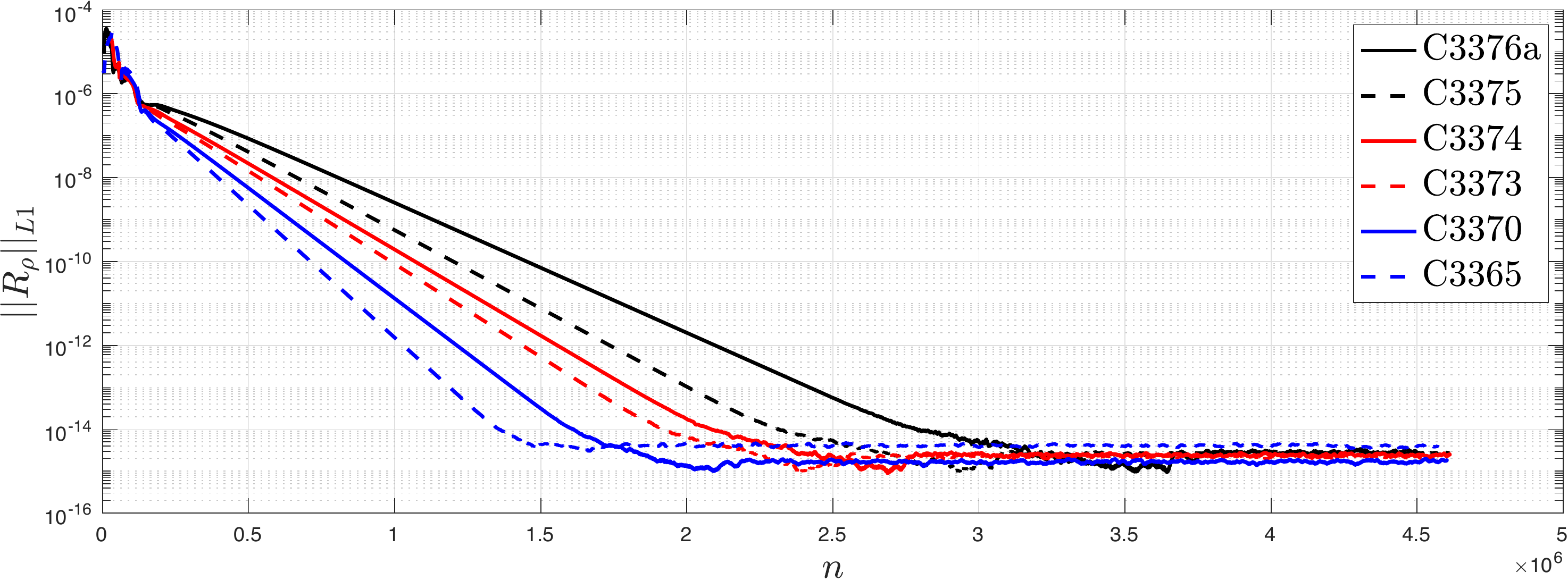}
\end{center}
\caption{Converging history of the base flow calculations with high-order shock-fitting method. The vertical axis represents the maximum residual \(||R_{\rho}||_{L_1}\) in density \(\rho\).}
\label{Fig2}
\end{figure}

The evolution of the maximum density residual as a function of the number of time steps is shown in figure \ref{Fig2}. The small initial residual level is due to the well-converged initial field from the preliminary calculation using first-order upwind scheme. After several millions of steps, when the residual reaches the machine accuracy, this ``steady-state" is considered as converged. From the figure \ref{Fig2}, one can find that  cases with higher Reynolds number \(Re_{\infty}\) converge slower. More time steps are obviously needed by the flow to adjust to the much thicker boundary layers where viscous effects are stronger.

\begin{figure}
\begin{center}
 \includegraphics[width=0.5\textwidth]{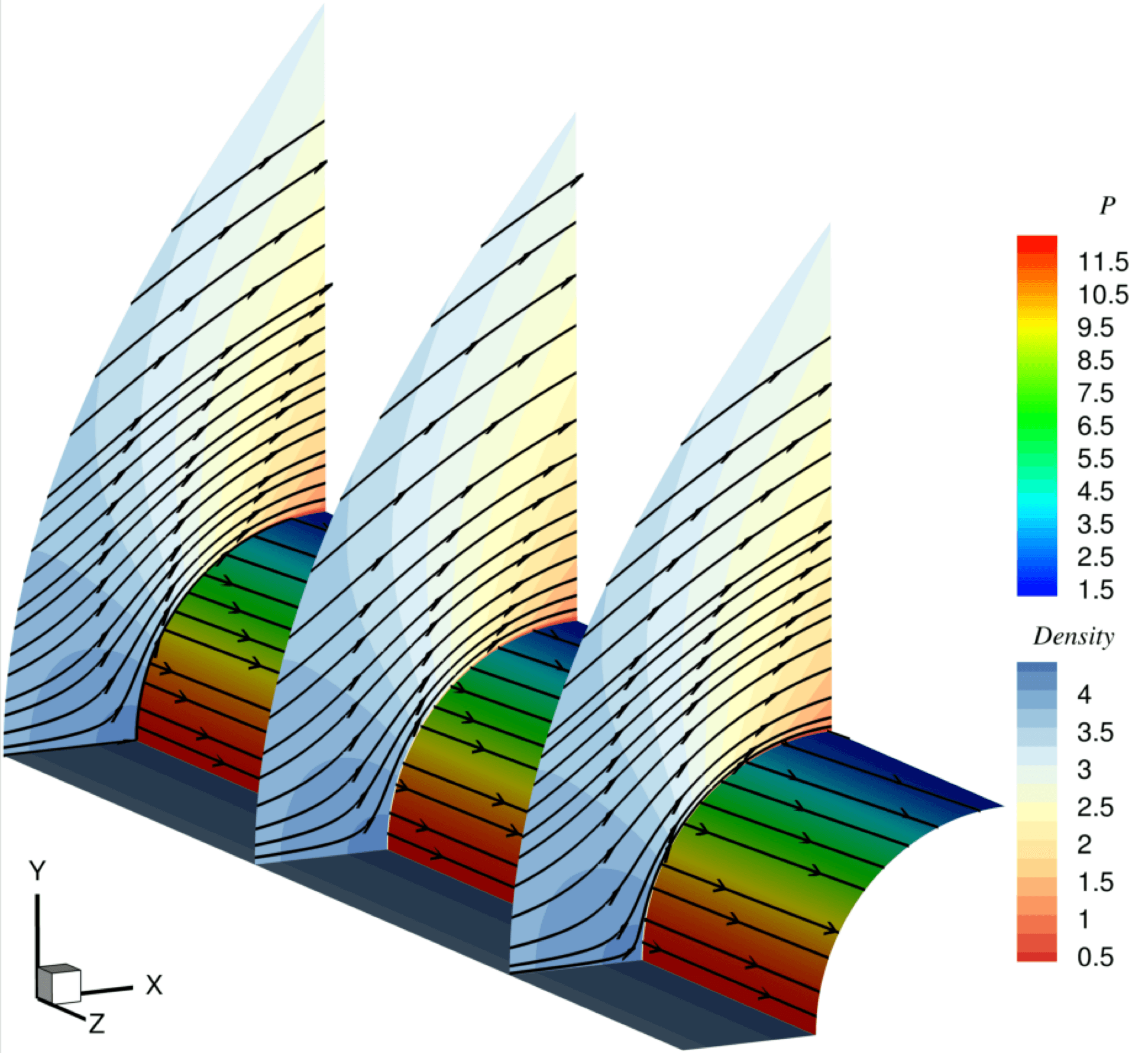}
\end{center}
\caption{Contours of base-flow density at three spanwise locations together with pressure contour over cylinder wall surface. Streamlines are also plotted on these contours.}
\label{Fig3}
\end{figure}

\begin{figure}
\begin{center}
\includegraphics[width=0.75\textwidth]{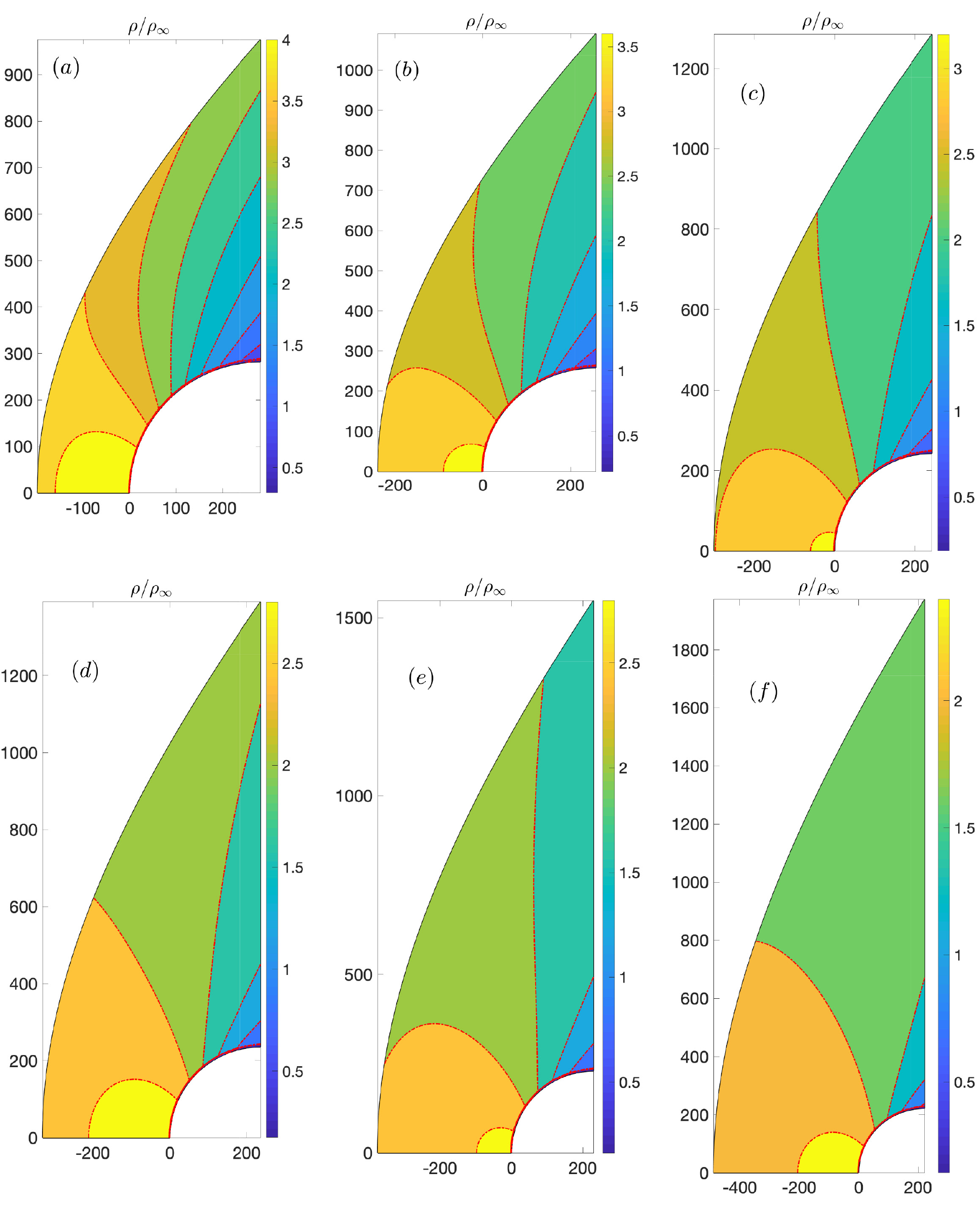} 
\end{center}
\caption{ Density contours over \(x-y\) plane for all cases: \((a)-(f)\) represent the cases from C3365 with sweep Mach number $M_s= 3.94$ to C3376a of $M_s=5.8$. Only upper half plane is shown because of symmetry.}
\label{Fig4}
\end{figure}

\begin{figure}
\begin{center}
\begin{tabular}{cc}
\includegraphics[width=0.45\textwidth]{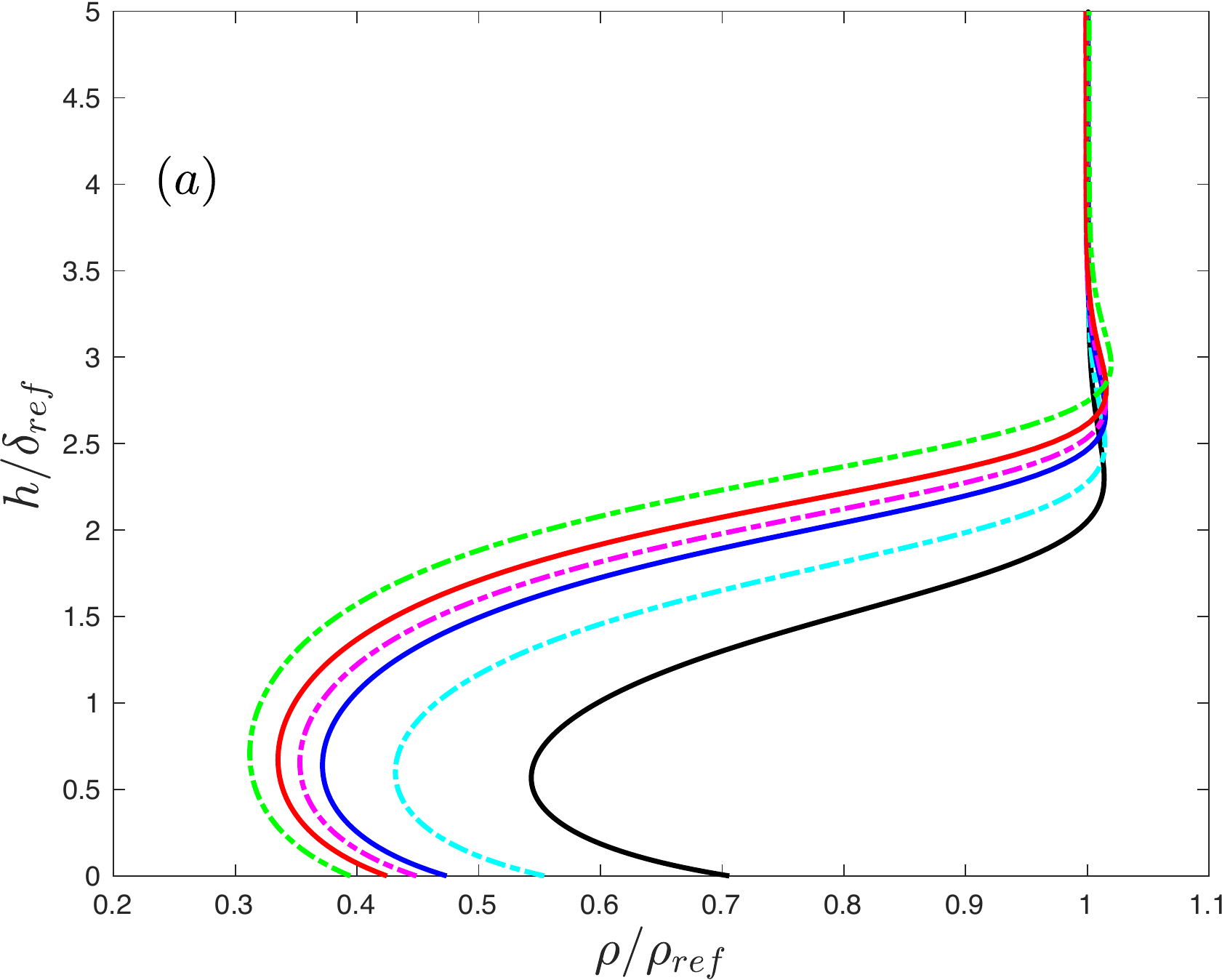} & \includegraphics[width=0.45\textwidth]{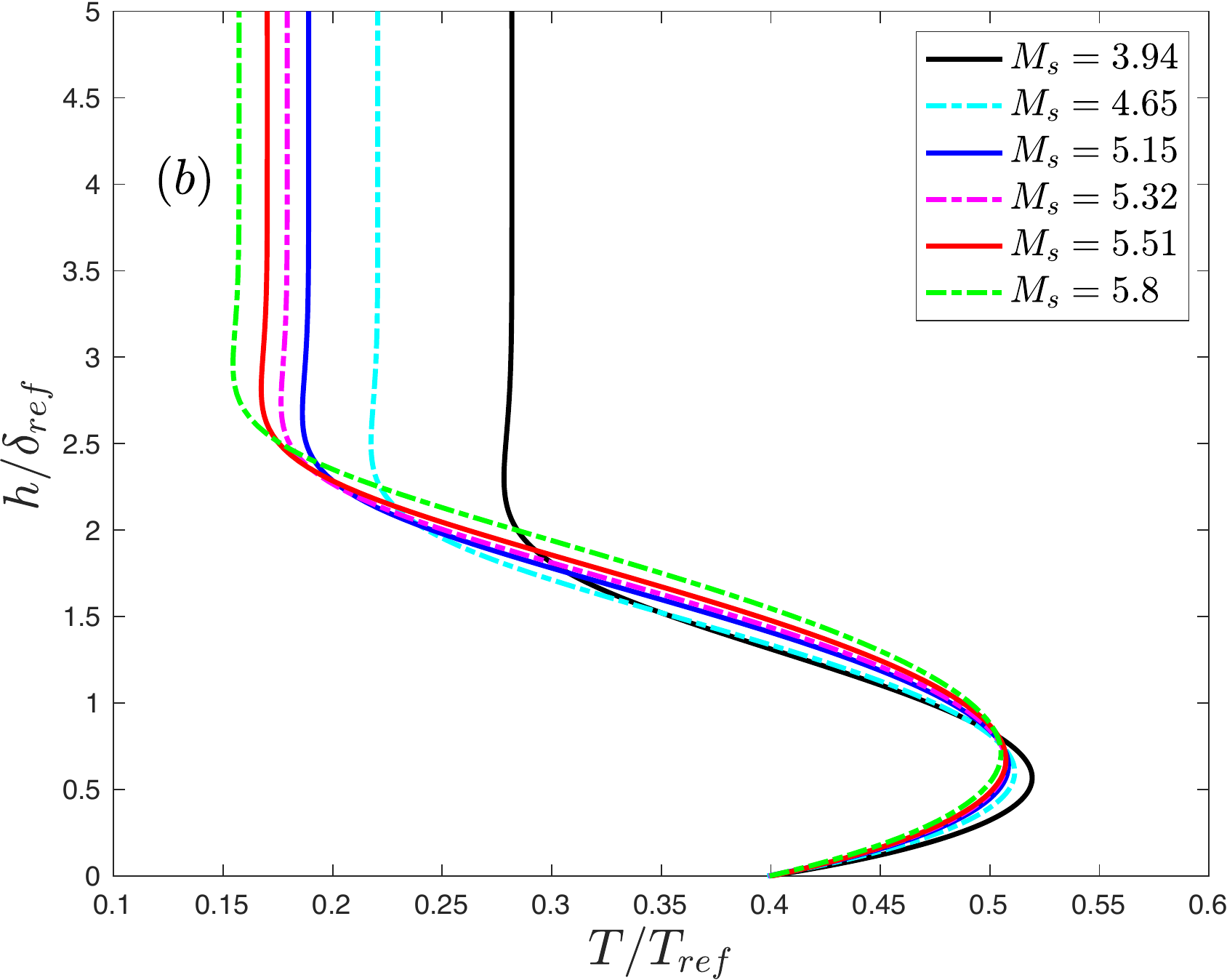} \\
\includegraphics[width=0.45\textwidth]{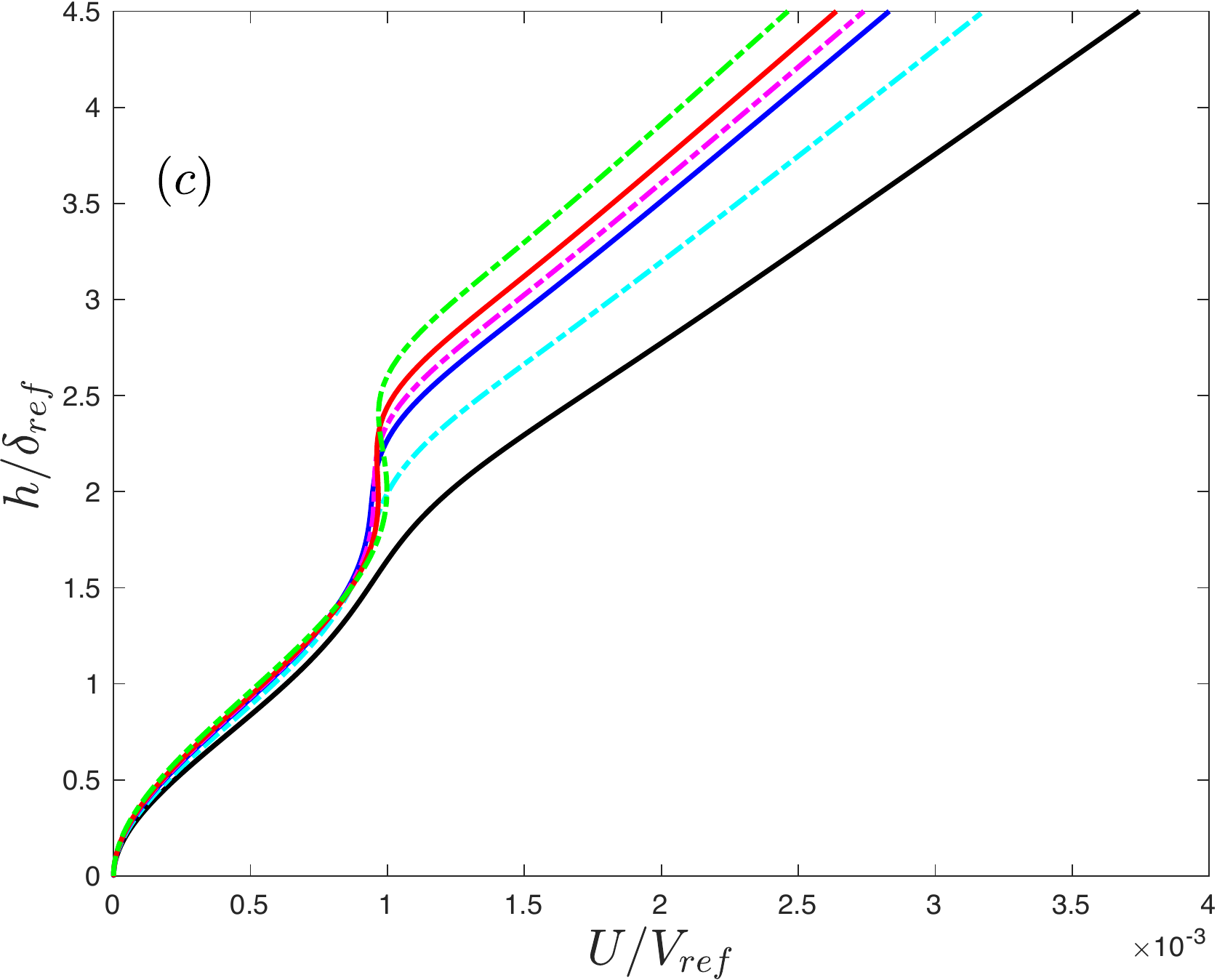} & \includegraphics[width=0.45\textwidth]{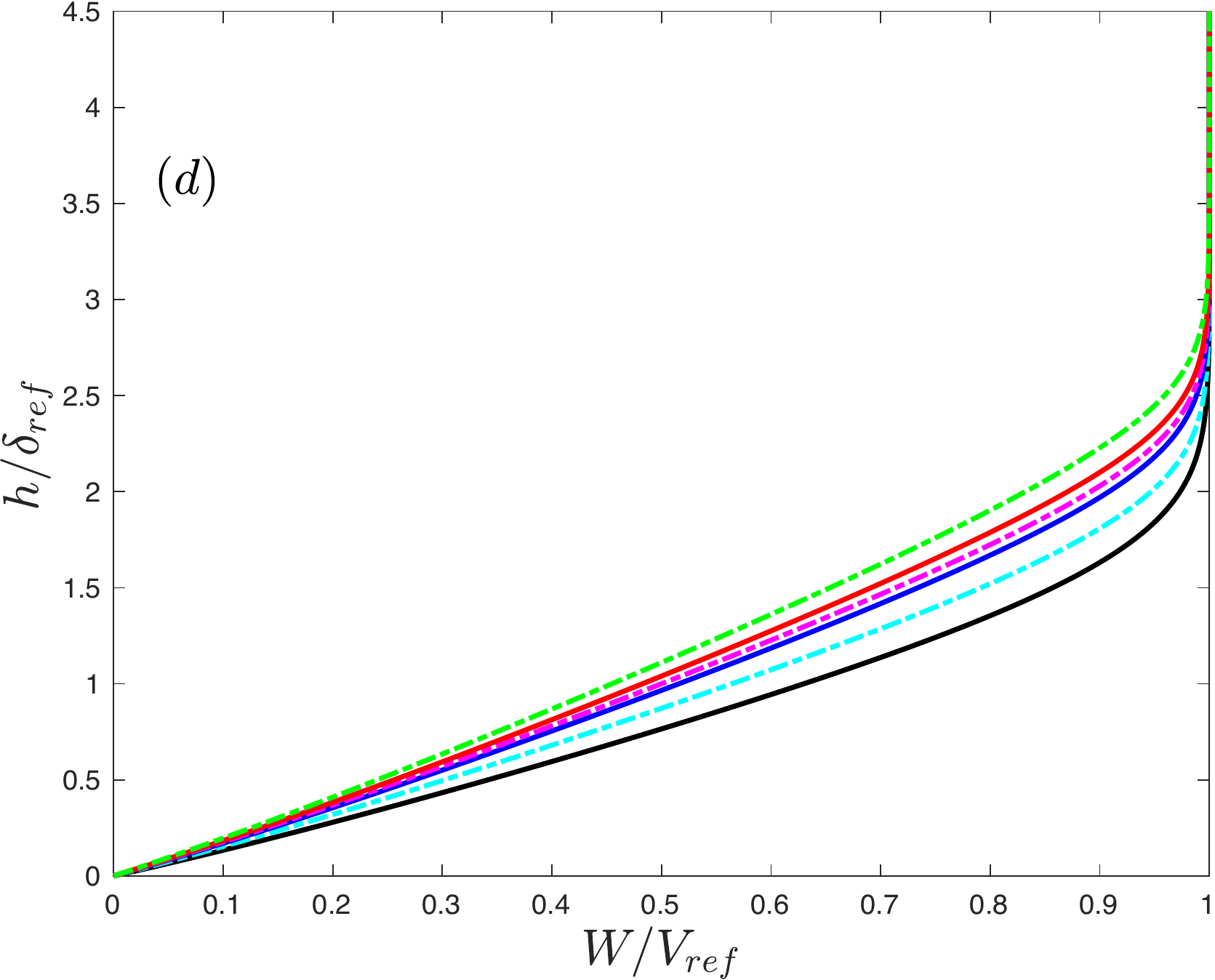}
\end{tabular}
\end{center}
\caption{Variation of the base-flow profiles with different sweep Mach numbers. Figures (a)-(d) represent the \(\rho,T\), \(U\) and \(W\) profiles, respectively. All reference values are defined at the edge of stagnation boundary layer except for temperature. The reference temperature takes the recovery temperature. \(\delta_{ref} = \delta\) and \(h\) represents the distance away from the attachment line.}
\label{Fig5}
\end{figure}

\begin{figure}
\begin{center}
\begin{tabular}{c}
\begin{overpic}
[scale=0.40]{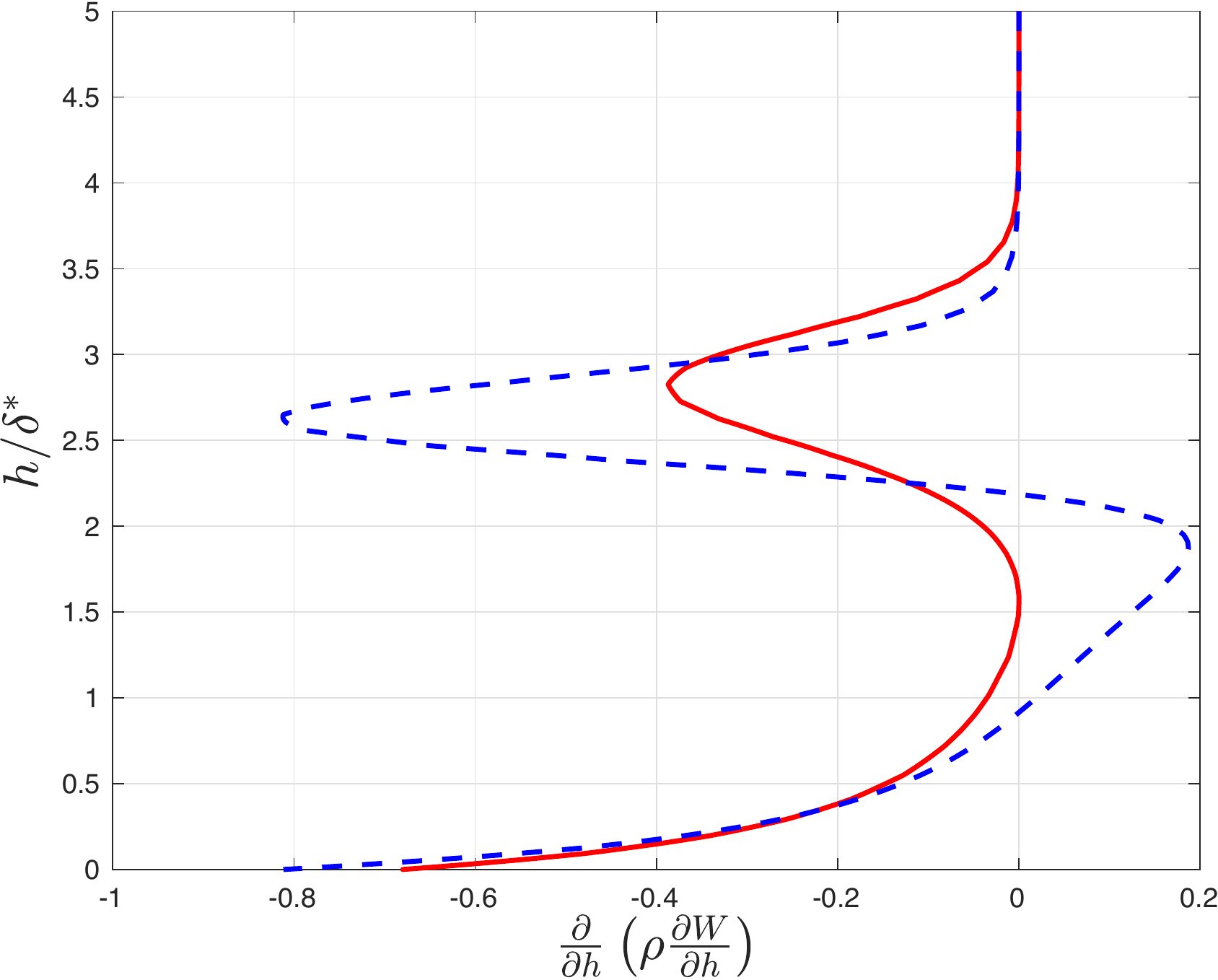}
\put(20,52){NS solution}
\put(70,25){BL}
\end{overpic}
\end{tabular}
\end{center}
\caption{The profiles of  \(\frac{\partial}{\partial h}\left( \rho\frac{\partial W}{\partial h} \right)\) along wall normal distance \(h/\delta^*\) from the attachment line for C3376a case with $ M_s=5.8$. The solid red line represents the result from the boundary-layer approximation and the dashed blue line the result from the full Navier-Stokes equation.}
\label{Fig5p}
\end{figure}

The flow field of C3365 case is visualized in figure \ref{Fig3} to illustrate key feature of the flow cases. As it can be observed, the curved streamlines in the \(x-y\) cross-plane around the cylinder together with the large spanwise velocity,  represent a typical three dimensional flow, especially at the leading edge.  The density distributions of the base flow for all cases are shown in figure \ref{Fig4}. At the leading edge, as the sweep angle becomes large, the shock is moving away from the wall surface and the shock standoff distance, the distance between the shock surface and attachment-line, increases from around 0.7R to 2.2R. The profiles of the main physical components of the attachment-line boundary layer are shown in figure \ref{Fig5}. Two major features should be noticed. First, as sweep Mach numbers increase, the thickness of the boundary layer increases. Second, interestingly, in the profiles of the \(U\)-velocity component, a distinct contortion is observed near the outer edge of the boundary layer. Also there, the temperature \(T\) and density \(\rho\) profiles exhibit variations which were not found in the solution of the boundary layer equations (see Appendix \ref{appC}). 
The base flow obtained with traditional boundary-layer assumptions is given in Appendix \ref{appC}. It will be seen there that the differences in the base flows are significant giving rise to the findings of the attachment-line modes.
The profile of \(\frac{\partial}{\partial h}\left(\rho \frac{\partial W}{\partial h}\right)\) at attachment line from C3376a case is shown in figure \ref{Fig5p}. By comparing profiles from the boundary-layer approximations and the full NS solution, the major differences between these two solutions can be easily found and two generalized inflection points, where \(\frac{\partial}{\partial h}\left(\rho \frac{\partial W}{\partial h}\right)=0\), are seen in figure \ref{Fig5p} in the NS solution. 

\begin{figure}
\begin{center}
\begin{tabular}{cc}
\includegraphics[width=0.45\textwidth]{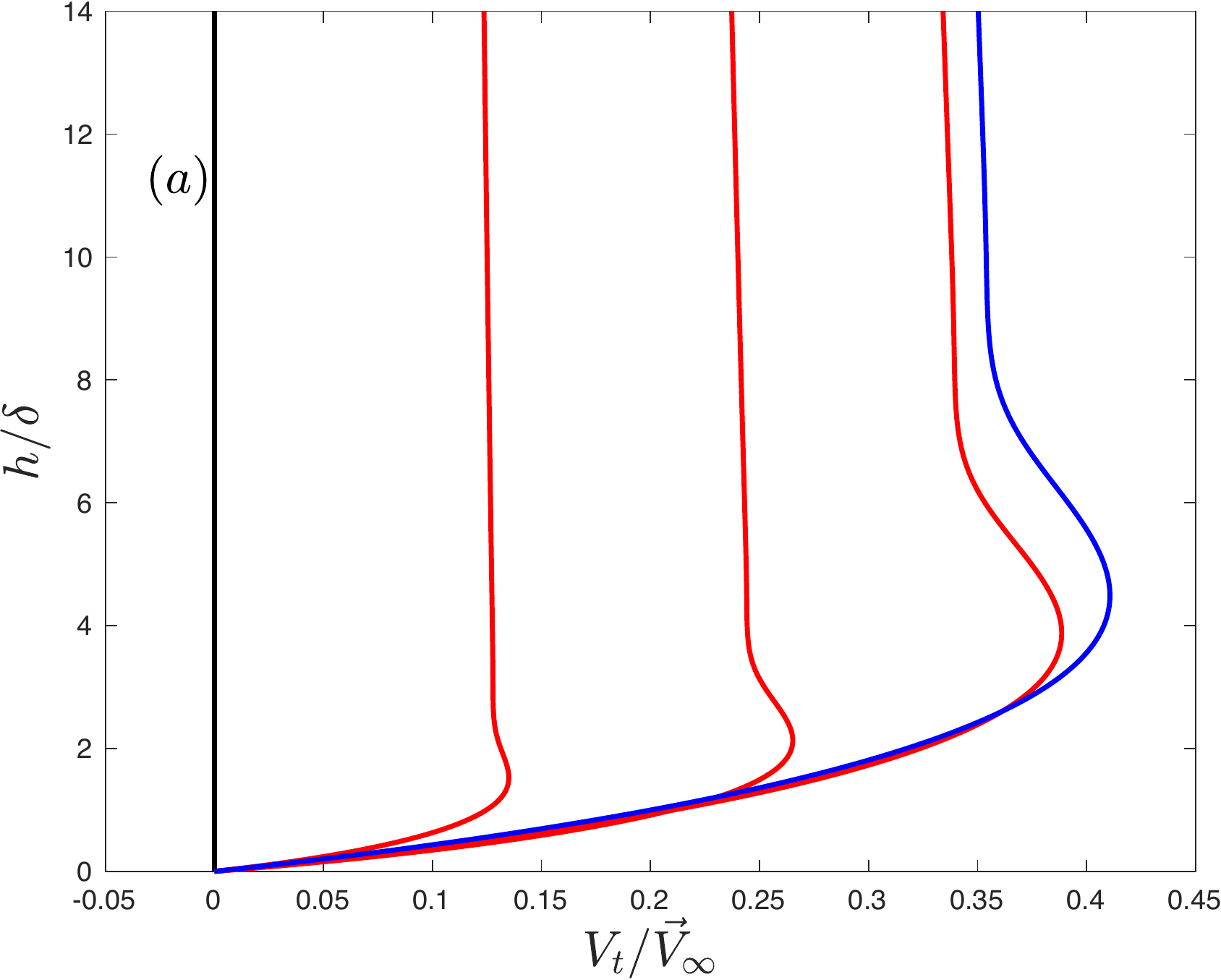} & \includegraphics[width=0.45\textwidth]{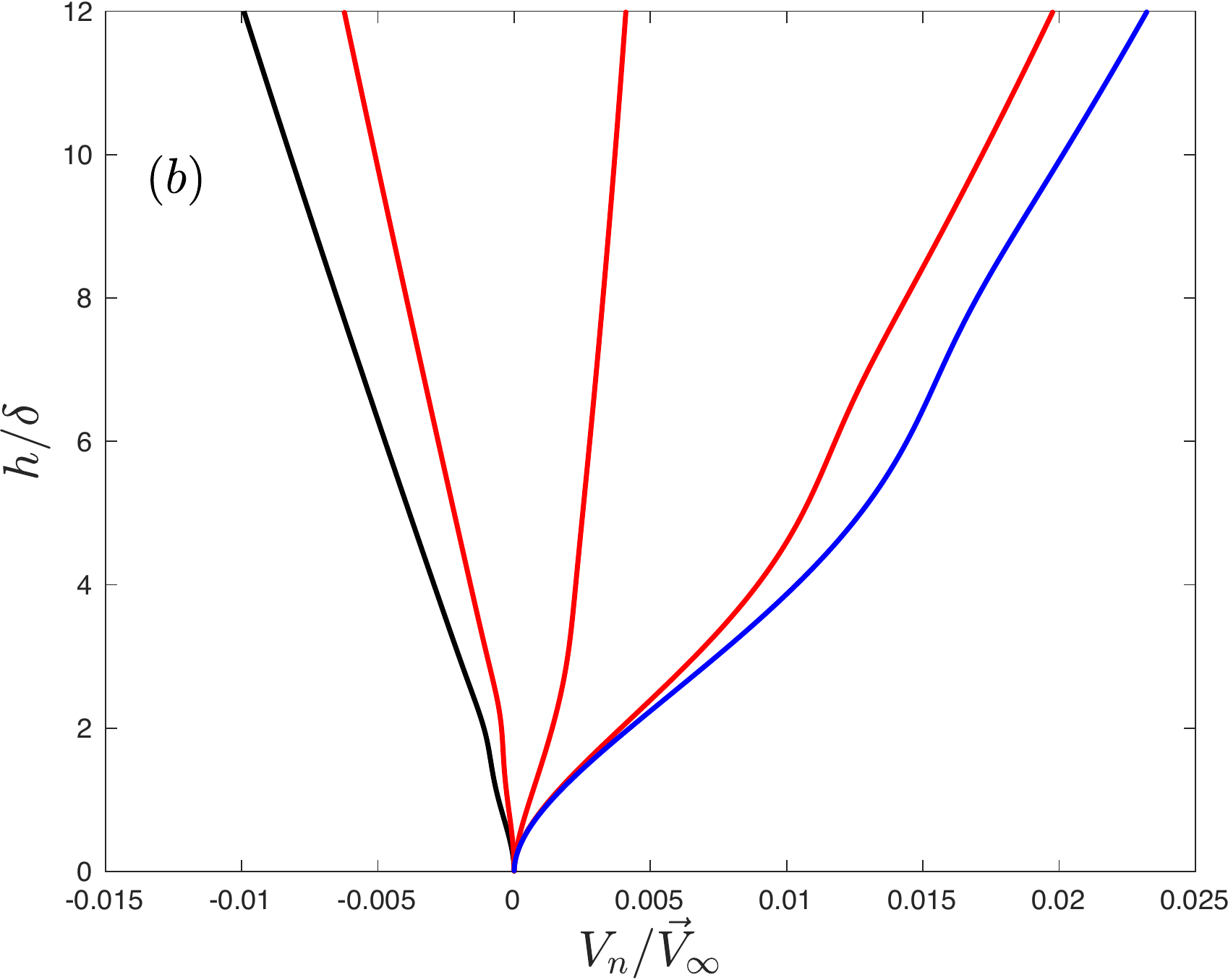} \\
\includegraphics[width=0.45\textwidth]{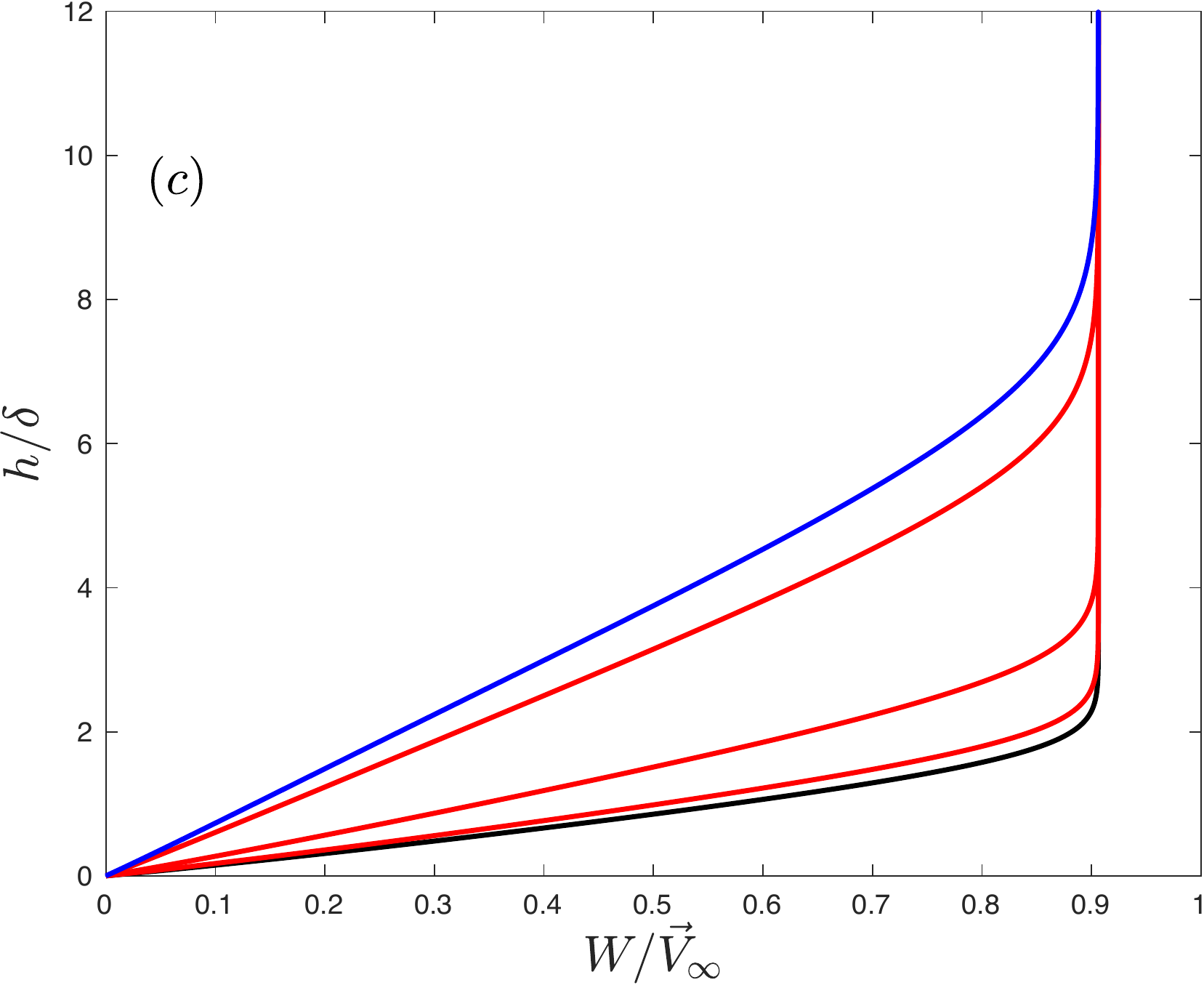} & \includegraphics[width=0.45\textwidth]{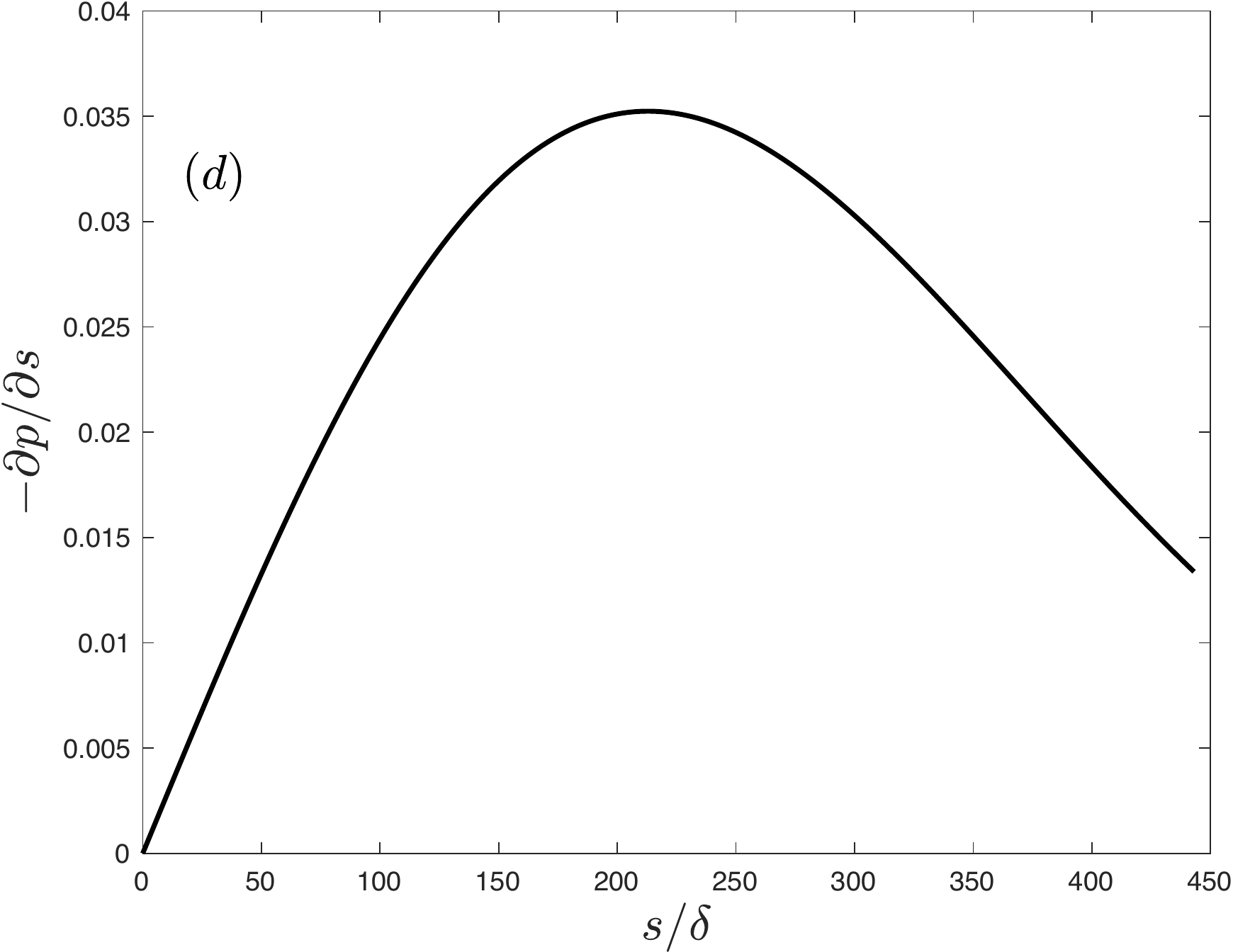}
\end{tabular}
\end{center}
\caption{Profiles of velocity components and pressure gradient in the surface direction for case C3365. Figure (a)-(c) show the tangential velocity \(V_t\), the normal velocity \(V_n\) and the sweep velocity \(W\) profiles along the surface from the attachment line (black lines, $s=0$) to the exit (blue lines, $s= 443.04$). Red lines, between the black and the blue, represent the velocity profiles at three increasing locations $s =$ 138.93, 277.86 and 416.79, respectively. The pressure gradient along the surface is shown in \((d)\). \(h\) represents the distance away from surface.}
\label{Fig6}
\end{figure}

Along the surface far from the attachment line, velocity profiles and pressure gradient at five different locations are shown in figure \ref{Fig6}. An inflection point appears along with the presence of tangential velocity overshoot in figure \ref{Fig6}(a), this is a typical phenomenon of a boundary layer with favorable pressure gradient. Along the surface, together with the development of the boundary layer, the spanwise profile becomes thicker (figure \ref{Fig6}(c)), and the wall-normal velocity profile turns from negative to positive(figure \ref{Fig6}(b)). The surface pressure gradient is also shown in figure \ref{Fig6}(d) and over the whole surface the fluid is accelerated continuously. 

\section{Stability analysis}\label{S4}

\begin{figure}
\begin{center}
\includegraphics[width=0.5\textwidth]{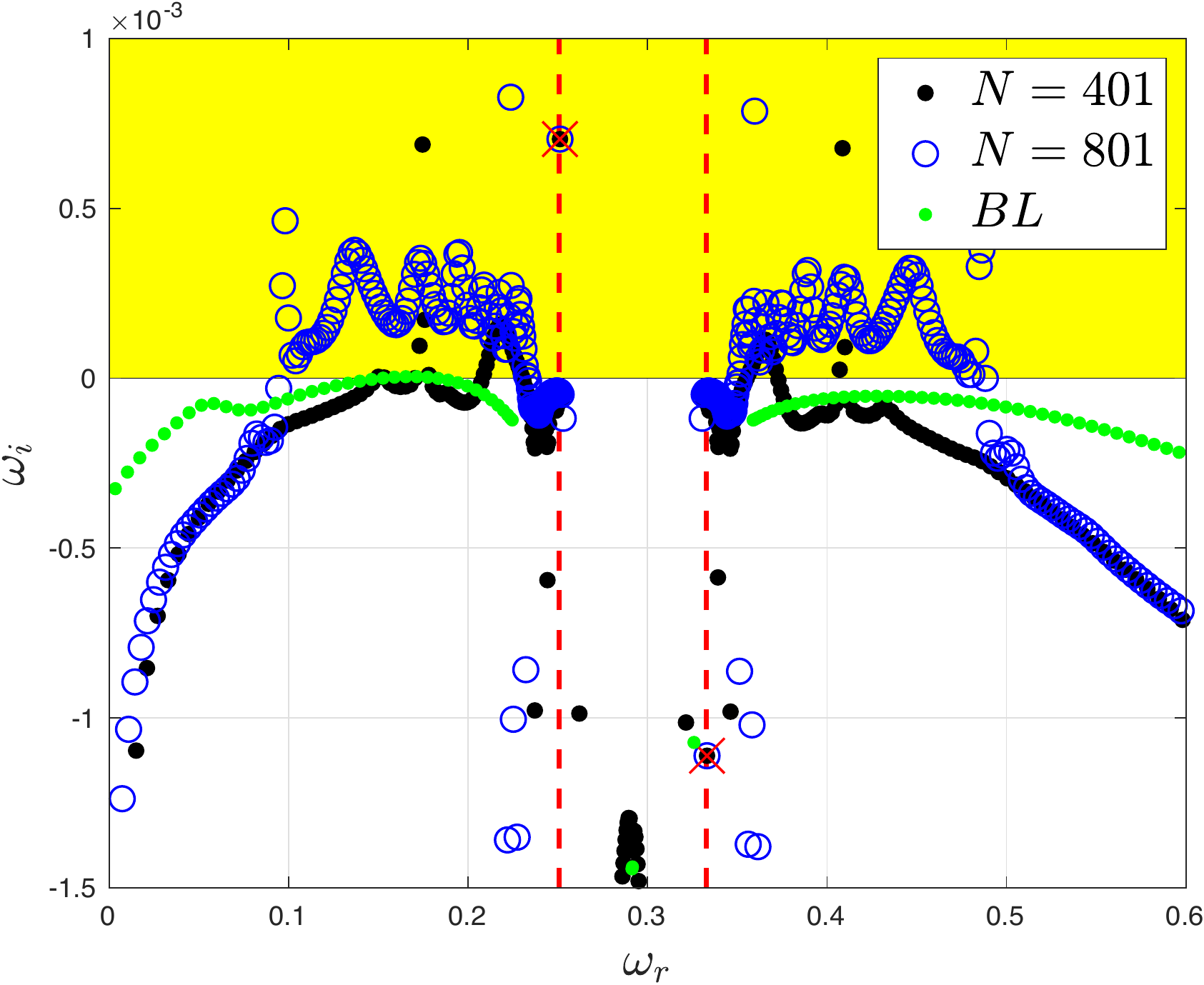}
\end{center}
\caption{Spectral distribution based on local analysis for C3376a case ($M_s=5.8$). The unstable region is marked in yellow. Two different grids had been used to cross-validate the results, and the discrete eigenvalues are marked by red cross. Two red dashed lines represent the locations of slow acoustic branch (the left one) and fast acoustic branch (the right one). The spectrum from the boundary-layer solution is shown by green points.}
\label{Fig7}
\end{figure}

In the present stability analysis, the behaviors of the perturbations at the attachment line are obtained both locally and globally. Firstly, the local analysis is performed along the attachment line based on the profiles from the previous full Navier-Stokes calculation.  Two sets of grids (401 and 801 points in the wall-normal direction, \(h\)), together with the spectral methods, had been employed to achieve the mesh-independent solution and to reveal the structure of the spectrum. Figure \ref{Fig7} shows the typical eigenspectrum of C3376a case, for illustration, based on the profiles from DNS calculation and the solution based on boundary layer approximation is also shown for comparison. Other cases have similar features. Two discrete modes are identified and marked in this figure and no unstable discrete mode is found when the base flow is calculated with boundary-layer equations. The unstable discrete mode locates around the continuous branch of the slow acoustic wave (the left red line). The stable one is found at around the fast acoustic wave (the right red line). The distribution of the spectrum is similar to the cases of hypersonic boundary layers over a cold wall \citep{Fedorov2011b}. However, because of the variations of base flow outside the boundary layer, the shape of the continuous spectrum changes significantly when more grids are used.

\begin{figure}
\begin{center}
\begin{tabular}{cc}
\includegraphics[width=0.45\textwidth]{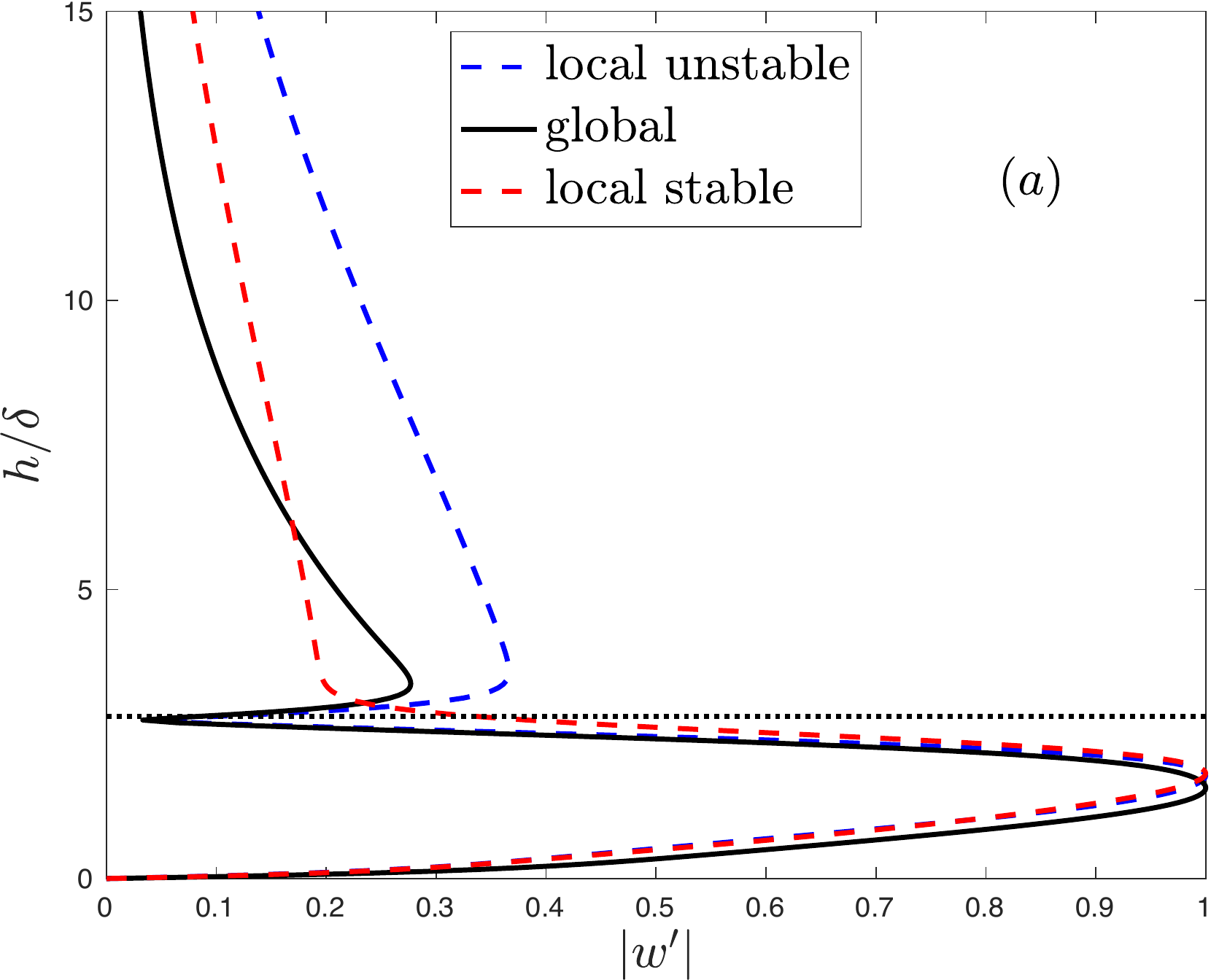} & \includegraphics[width=0.45\textwidth]{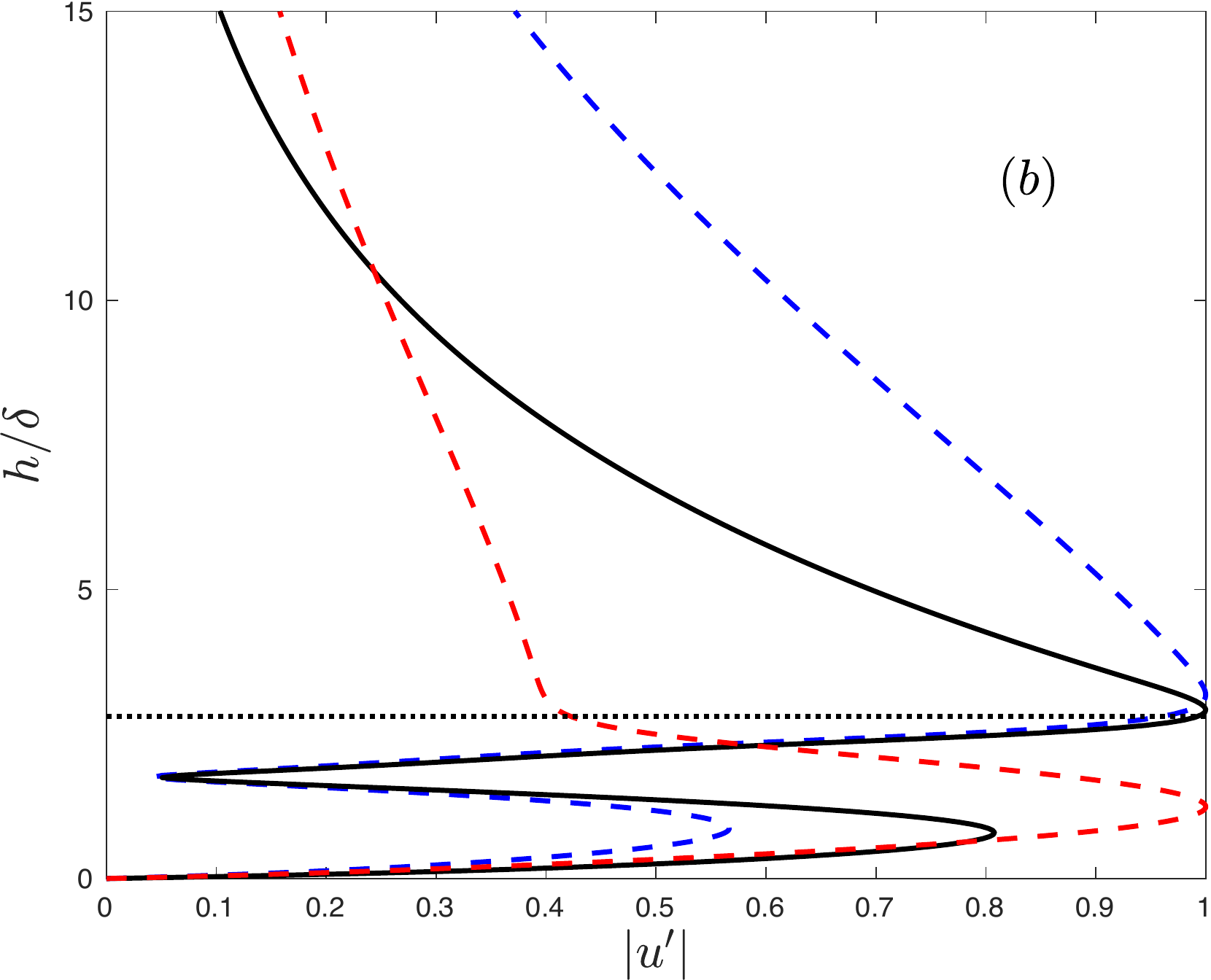} \\
\includegraphics[width=0.45\textwidth]{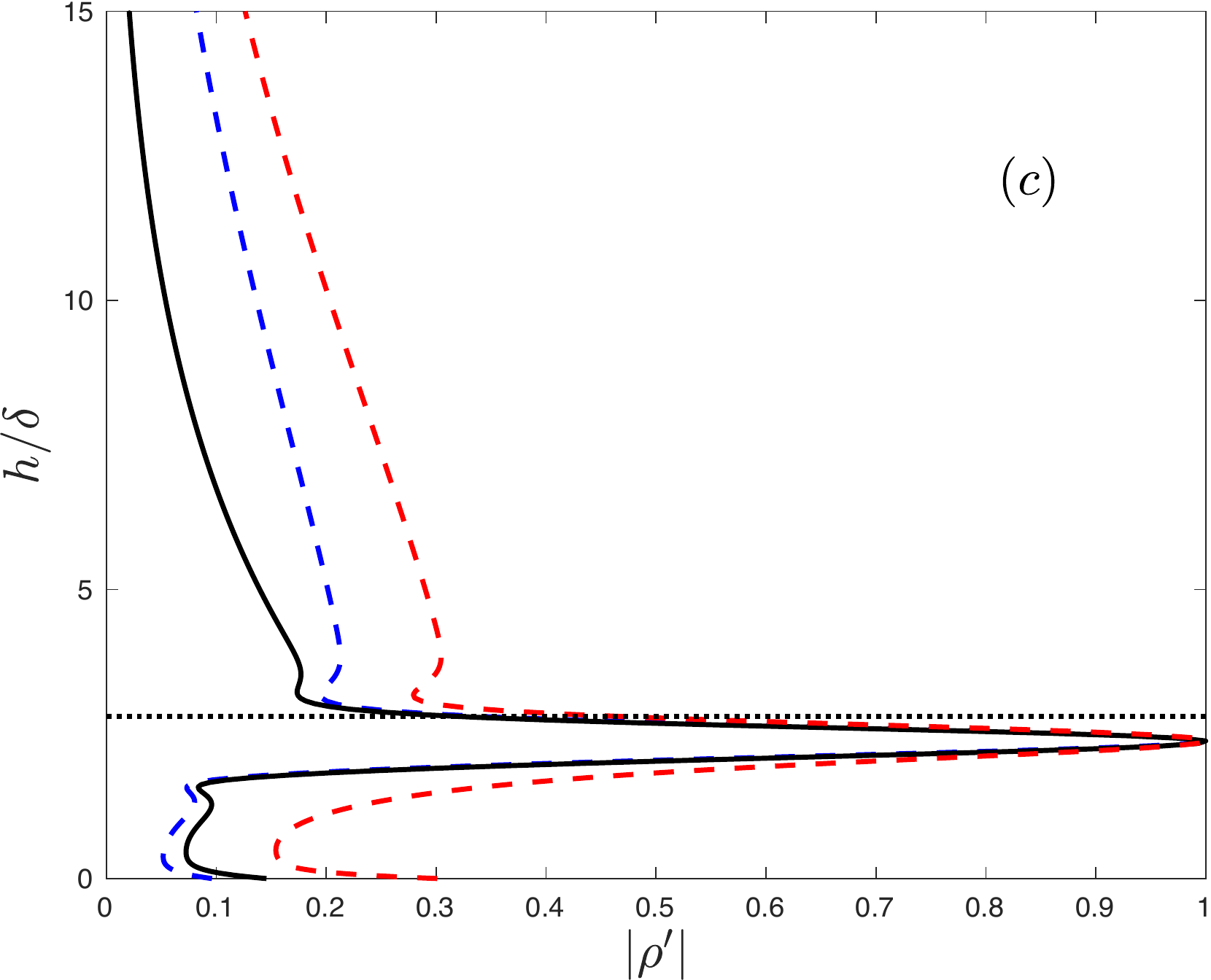} & \includegraphics[width=0.45\textwidth]{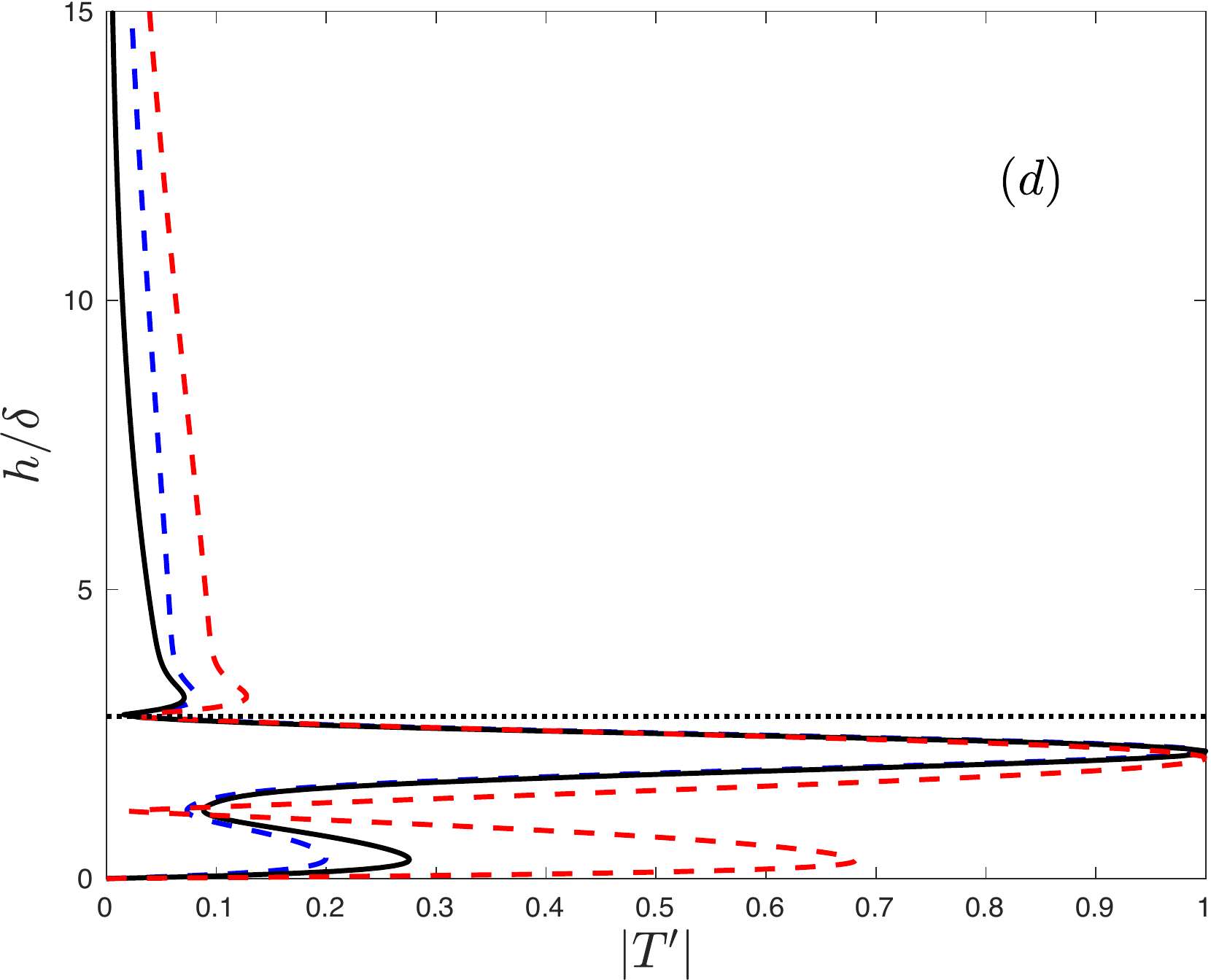} 
\end{tabular}
\end{center}
\caption{Comparisons of normalized perturbation profiles from the attachment line with solid black line from global stability analysis, dashed blue/red lines from local stability analysis for C3376a case ($M_s=5.8$). The dashed blue and red lines represent the eigenfunctions of unstable and stable discrete modes, respectively. All the eigenfunctions are normalized by the maximum norm with \((a)\) representing spanwise velocity perturbation \(|w'|\), \((b)\)  wall-normal velocity perturbation \(|u'|\), \((c)\) and \((d)\)  density and temperature perturbations. The black dotted lines represent the edge of the boundary layer.}
\label{Fig8}
\end{figure}

The eigenfunctions of this case are shown in figure \ref{Fig8} and the relative eigenvalue of unstable mode is shown in table \ref{table2}. All perturbations are normalized with their maximum norm. 
The perturbations are mainly distributed inside the boundary layer and become significant near the boundary layer edge. Outside of the boundary layer,  perturbations decay. For unstable modes, indicated as blue dashed lines and black lines, the results from the local calculation and the global calculation agree well. The amplitudes of the unstable eigenfunctions from the global calculation are larger than those from the local calculation inside the boundary layer, but decay much faster outside of the boundary layer, which can be seen in figure \ref{Fig8}\((b)\).

\begin{figure}
\begin{center}
\includegraphics[width=0.5\textwidth]{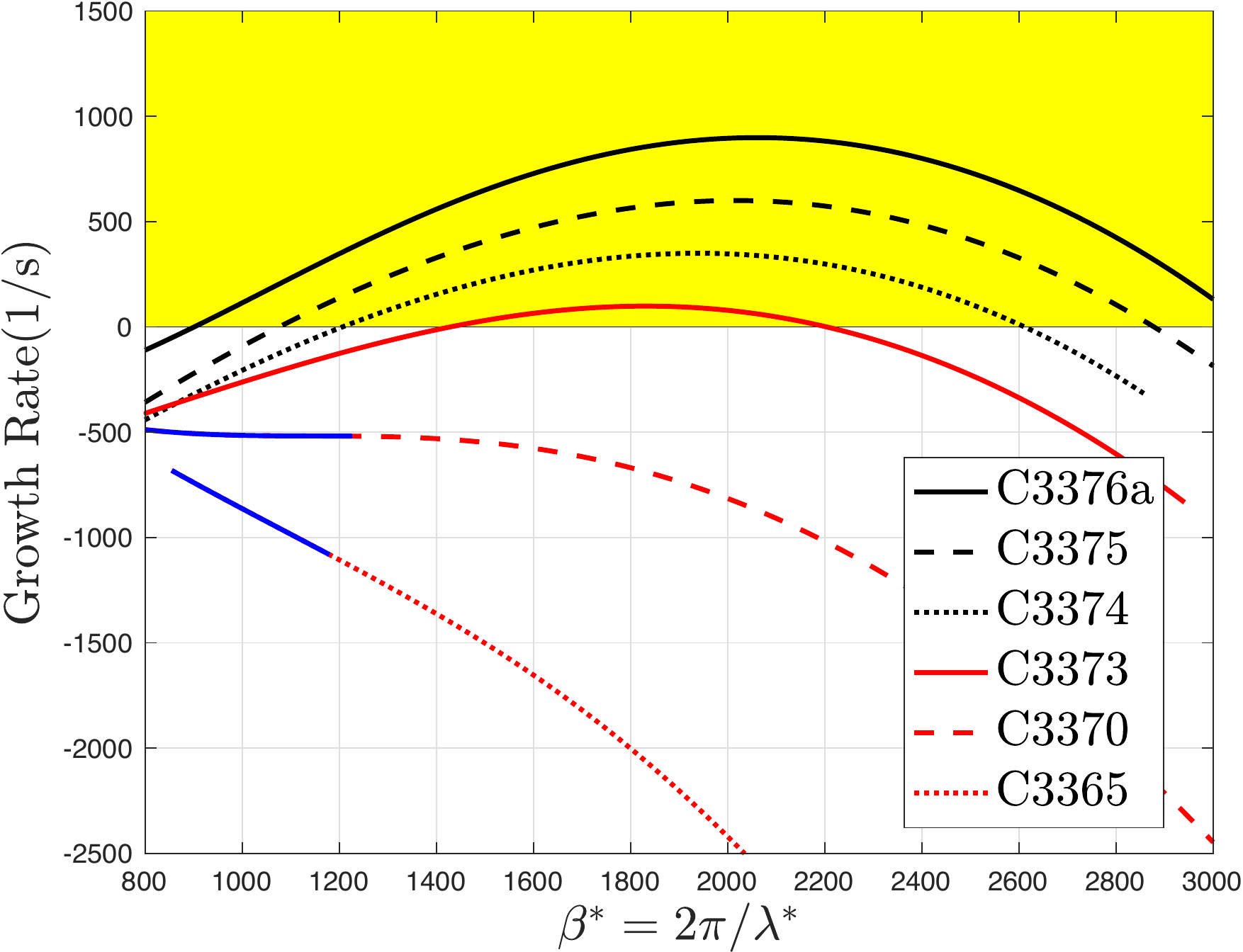}
\end{center}
\caption{Variations of growth rate of leading boundary modes with spanwise wave numbers for all cases. The blue lines represent regions where the discrete modes are absorbed into continuous branches. \(\lambda^*\) represent the dimensional wave length of the perturbations along \(z\) direction.}
\label{Fig9}
\end{figure}

In reality, physical perturbations consist of waves with various wave numbers. It is thus interesting to investigate the reliance of local growth rates to spanwise wave numbers. To also compare results among different cases, a dimensional spanwise wave number \(\beta^* = \beta / \delta\) is used. As shown in figure \ref{Fig9}, when the sweep Mach number increases from 3.94 in C3365 to 5.8 in C3376a, the unstable region is broadened and the local growth becomes larger. This finding is totally different from low-speed situation. For subsonic flow, as reported in \citet{Gennaro2013}, when the sweep Mach number decreases the growth rate increases. Moreover, for the cases with low spanwise Mach numbers,  3.94 in C3365 and 4.65 in C3370, the leading discrete modes are absorbed into continuous branches at small \(\beta^*\) and can not be tracked as shown by the blue lines in figure \ref{Fig9}.

\begin{figure}
\begin{center}
\includegraphics[width=0.5\textwidth]{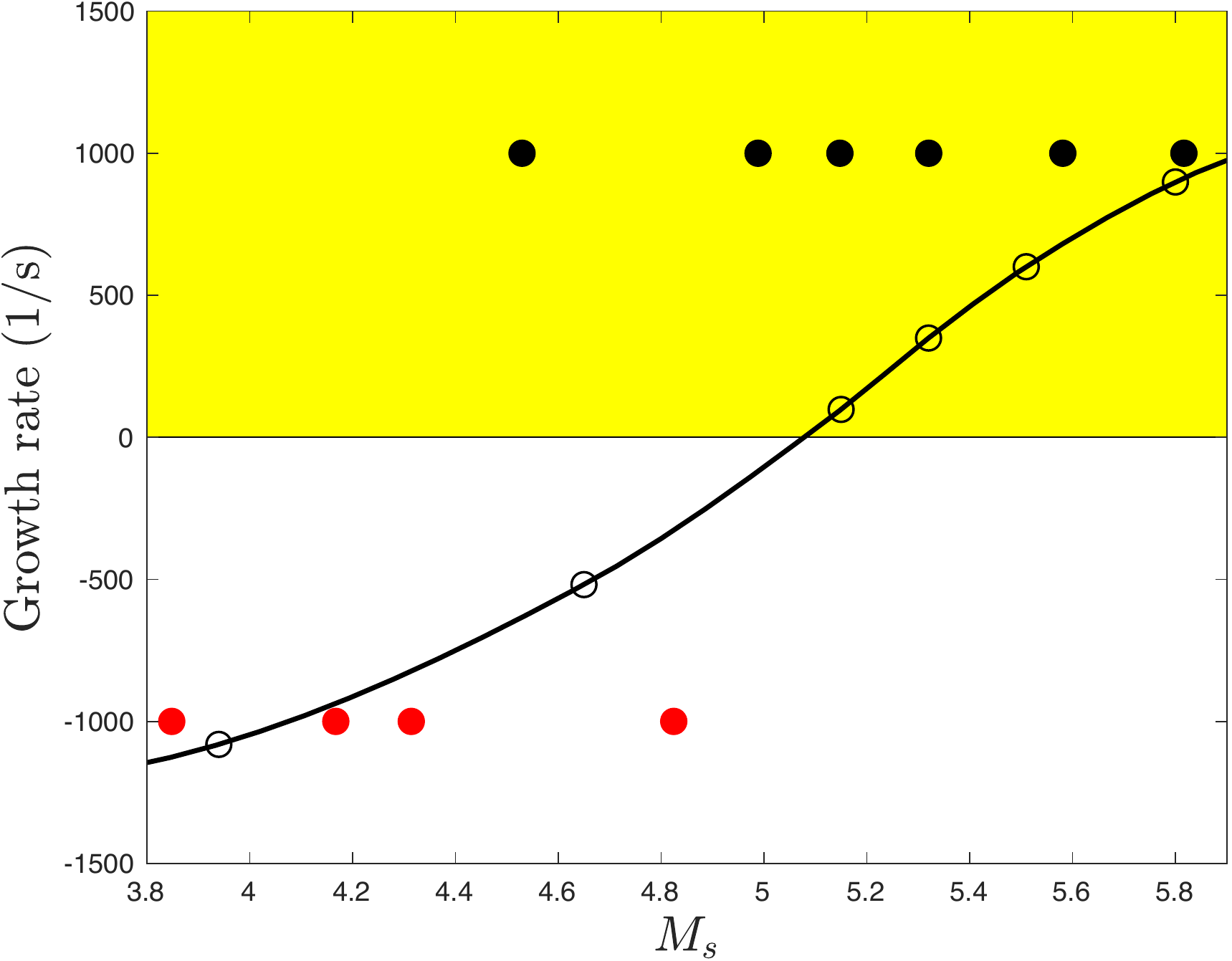}
\end{center}
\caption{Variation of the leading boundary modes with sweep Mach number. The solid black dots represent the cases where transition was detected at the attachment line in the experiments \citep{Gaillard1999} while the solid red dots indicate no transition. The black line with the circles is the result from the local analysis.}
\label{Fig10}
\end{figure}

Further comparison of the maximum growth rates of various sweep Mach numbers with the transition detections from experiments is shown in figure \ref{Fig10}. As reported in the experiment \citep{Gaillard1999}, when the sweep Mach number increases from around 3.5 to 6, the transition Reynolds number defined by \citet{Poll1979} decreases continuously. The theoretical growth rate increases continuously under similar conditions. In general, the behavior of these local modes agrees well with the experimental results. It explains why the critical transition Reynolds number decreases when the sweep Mach number is above 5. 
Together with the analysis of base flow (see figure \ref{Fig5p}), this attachment-line mode is different from incompressible cases (\citet{Lin1996,Lin1997}, \citet{Theofilis1995,Theofilis1998,Theofilis2003J} and \citet{Obrist2003a,Obrist2003b}). Traditional attachment-line modes for compressible flow can be treated as a kind of three-dimensional TS waves which belongs to viscous instability \citep{Lin1995}. Based on the velocity profiles at attachment line (figure \ref{Fig5}), the major base flow components along the line are the density, temperature and spanwise velocity. The velocity components in the \(x-y\) plane are a few orders smaller than that of the spanwise velocity and can be ignored from the leading term analysis (see Appendix \ref{appD}). Thus, the boundary layer along the attachment line can be seen as a parallel flow and is similar to the boundary layer along a flat plate. In fact, with the help of classical inviscid theory \citep{Lees1946,Mack1984}, the attachment-line mode found in this study belongs to the inviscid instability.

\begin{table}
\centering
\setlength{\tabcolsep}{0.04\textwidth}
\begin{tabular}{ccccc}
                           &\(N_s\)     &  \(N_n\)    & $\omega_r$ & $\omega_i$ \\
Local Calculation&                &   801        & 0.25120       & 0.00070372 \\
Global Calculation&  401       &   401        & 0.25097       & 0.00073089 \\      
Global Calculation&  601       &   401        & 0.25097       & 0.00073146 \\                   
\end{tabular}
\caption{Comparison of the local stability result together with the global results for the case C3376a with sweep Mach number $M_s=5.8$. \(N_s\) and \(N_n\) represent the grid points along surface and wall normal direction, respectively.}
\label{table2}
\end{table}

The major limitation in local stability theory is the neglection of the multi-dimensional effect which can be easily identified in the base flow (figure \ref{Fig3}). In particular, in the vicinity of the attachment line, flow impingement rather than shear is the dominant feature. 
On the contrary, non-negligible variations of base flow with respect to \(y\) direction, the curvature effects around the attachment line and the features of further downstream region can all be taken into account properly by the global stability analysis. 

\begin{figure}
\begin{center}
\includegraphics[width=0.5\textwidth]{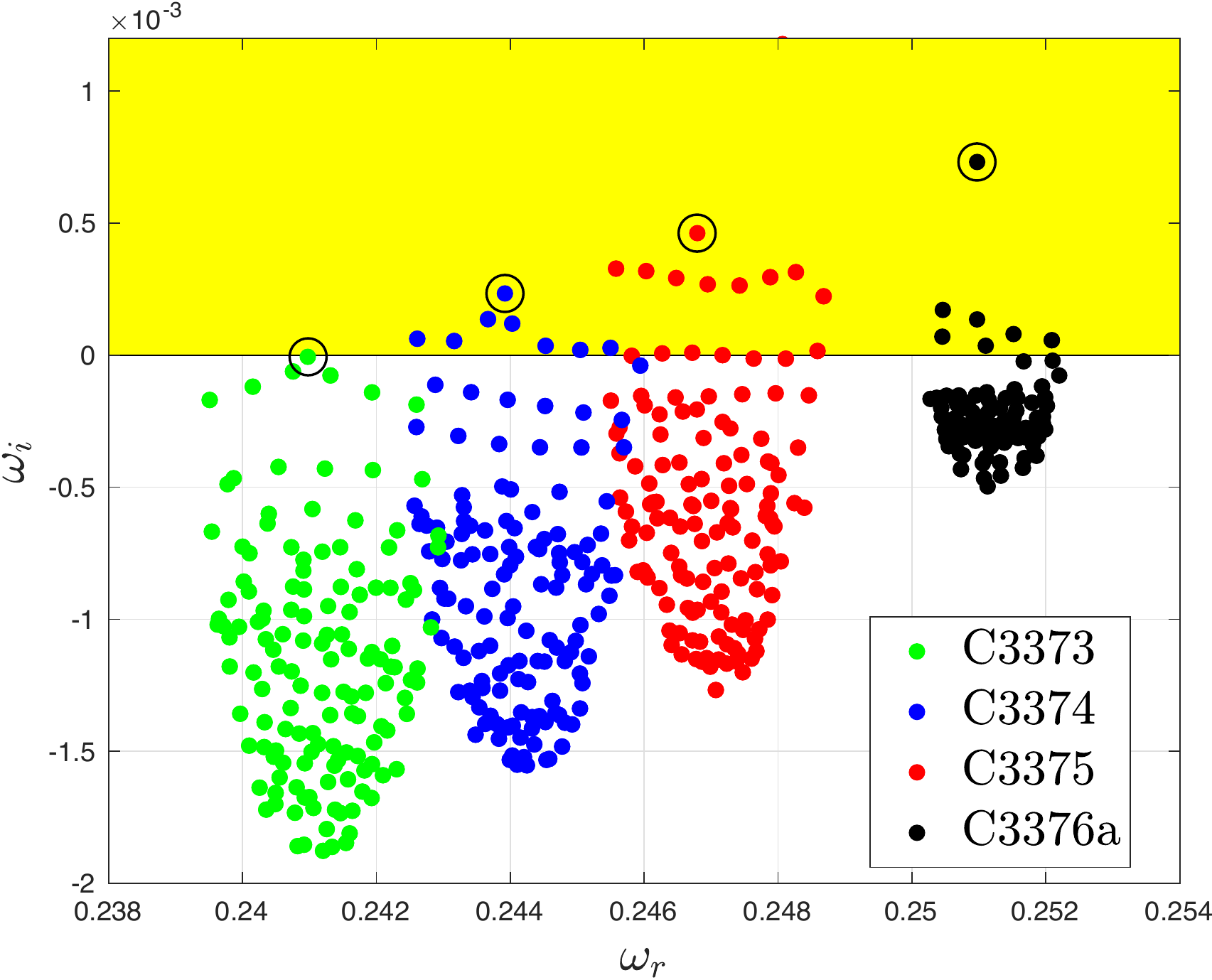}
\end{center}
\caption{The calculated spectrum of the unstable eigenspectrum for sweep Mach numbers 5.8 (C3376a), 5.51 (C3375), 5.32 (C3374) and 5.15 (C3373). The leading eigenvalues are marked by black circles.}
\label{Fig11}
\end{figure}

The global instabilities are performed on a very fine FD-q grids with 601 grid points along the surface tangential direction \(s\) and \(401\) grid points on the wall normal direction \(n\) over the \(x-y\) cross-section plane. Compared with the results from lesser grids (as shown in table \ref{table2}), this resolution (\(601\times 401\)) can well capture the main feature of the global instabilities. The calculated eigenspectrum are shown in figure \ref{Fig11} for the four most dangerous cases at sweep Mach number greater than 5.  

\begin{figure}
\begin{center}
\includegraphics[width=0.60\textwidth]{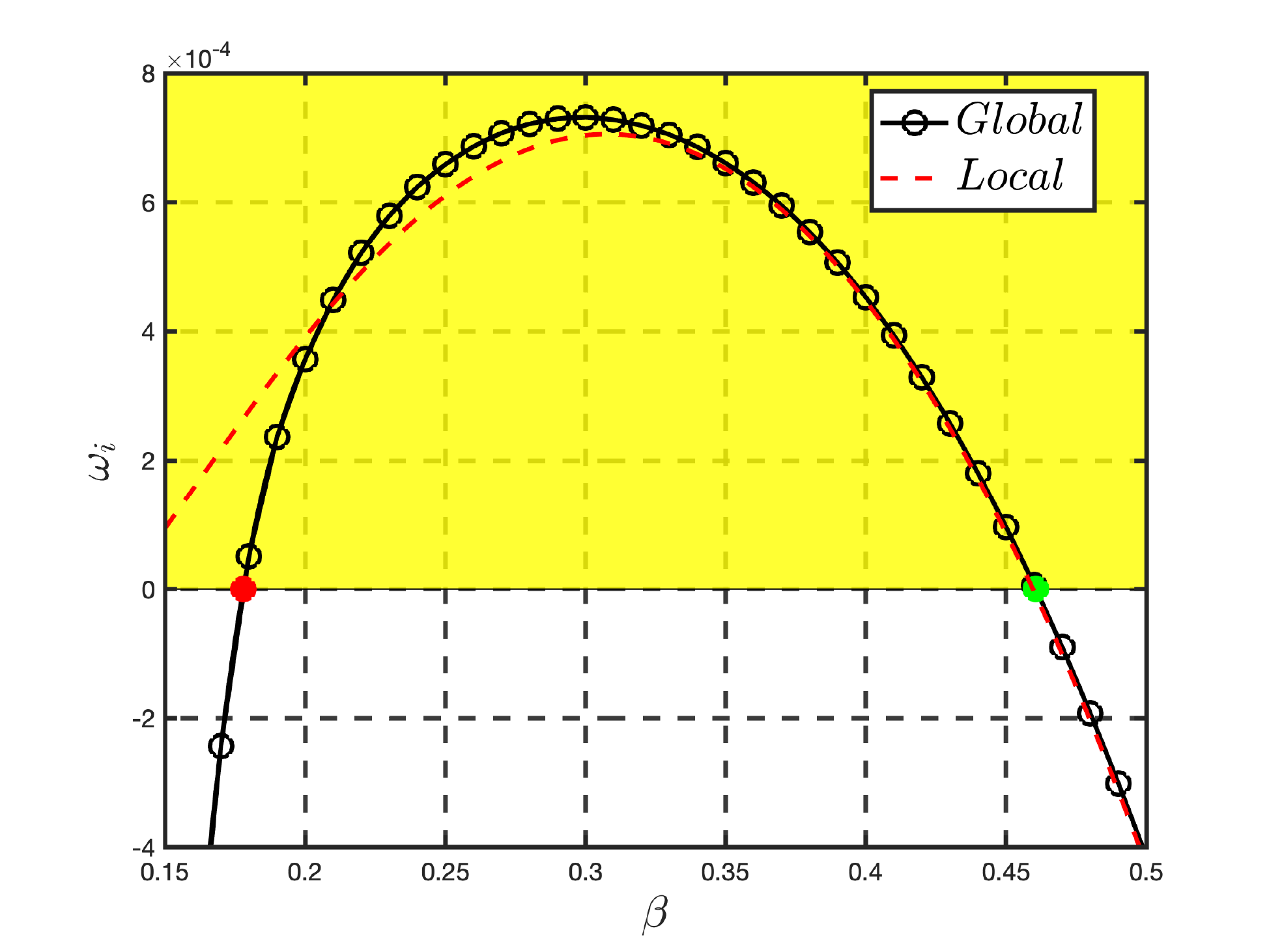}
\end{center}
\caption{Dependences of $\omega_i$ on the spanwise wave number $\beta$ for C3376a case with $M_s=5.8$. The black line with black circles represents the results from global calculations and the red dashed line represents the results from local calculation. The red and green dots represent two critical values.}
\label{Fig12}
\end{figure}

The dependence of $\omega_i$ on the spanwise wave number  $\beta$ is shown in figure \ref{Fig12} for both local and global calculations. It is seen here that the results from these two analyses agree reasonably well at wave number roughly greater than 0.2084. Less than this value the global growth rate drops much faster than local calculations. In fact, 
the global calculation indicates that the mode is unstable in the region \(0.178 <\beta< 0.461\). The maximum global growth rate is slightly larger than the local analysis. The major difference of local and global analysis for this case is the leading edge effect of the stability equation, flow impingement and curvature effects of the base flow are included in the solution of NS equations. Thus, for small spanwise wave number \(\beta\) the leading edge curvature has a stabilizing effect but a destabilizing effect when the wave number is larger in the unstable region. This finding is different from the results for incompressible flows where the leading edge curvature exhibits a stabilizing effect on the attachment-line boundary layer \citep{Lin1997}. 

\begin{figure}
\begin{center}
\begin{tabular}{cc}
\includegraphics[width=0.45\textwidth]{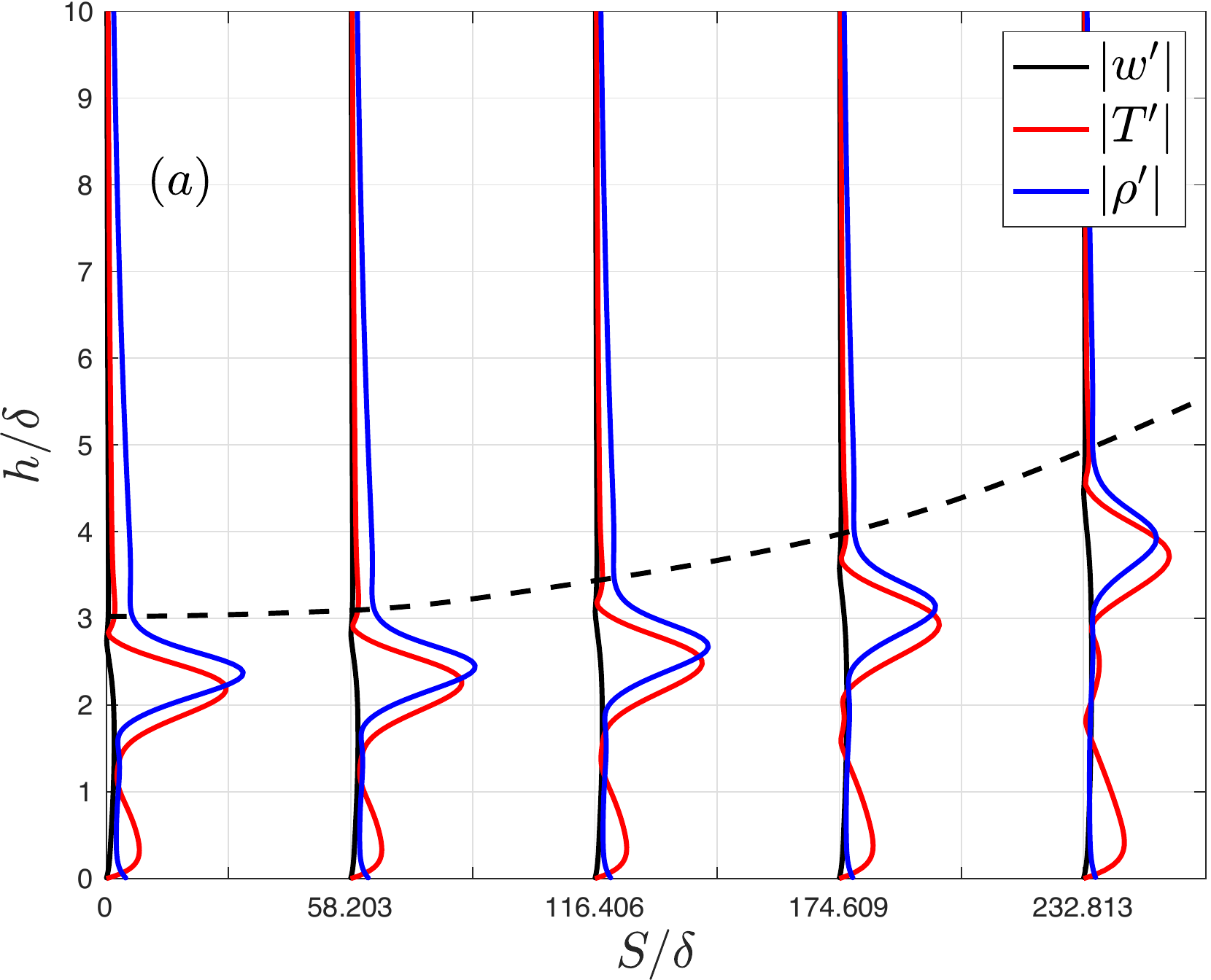} &
\includegraphics[width=0.45\textwidth]{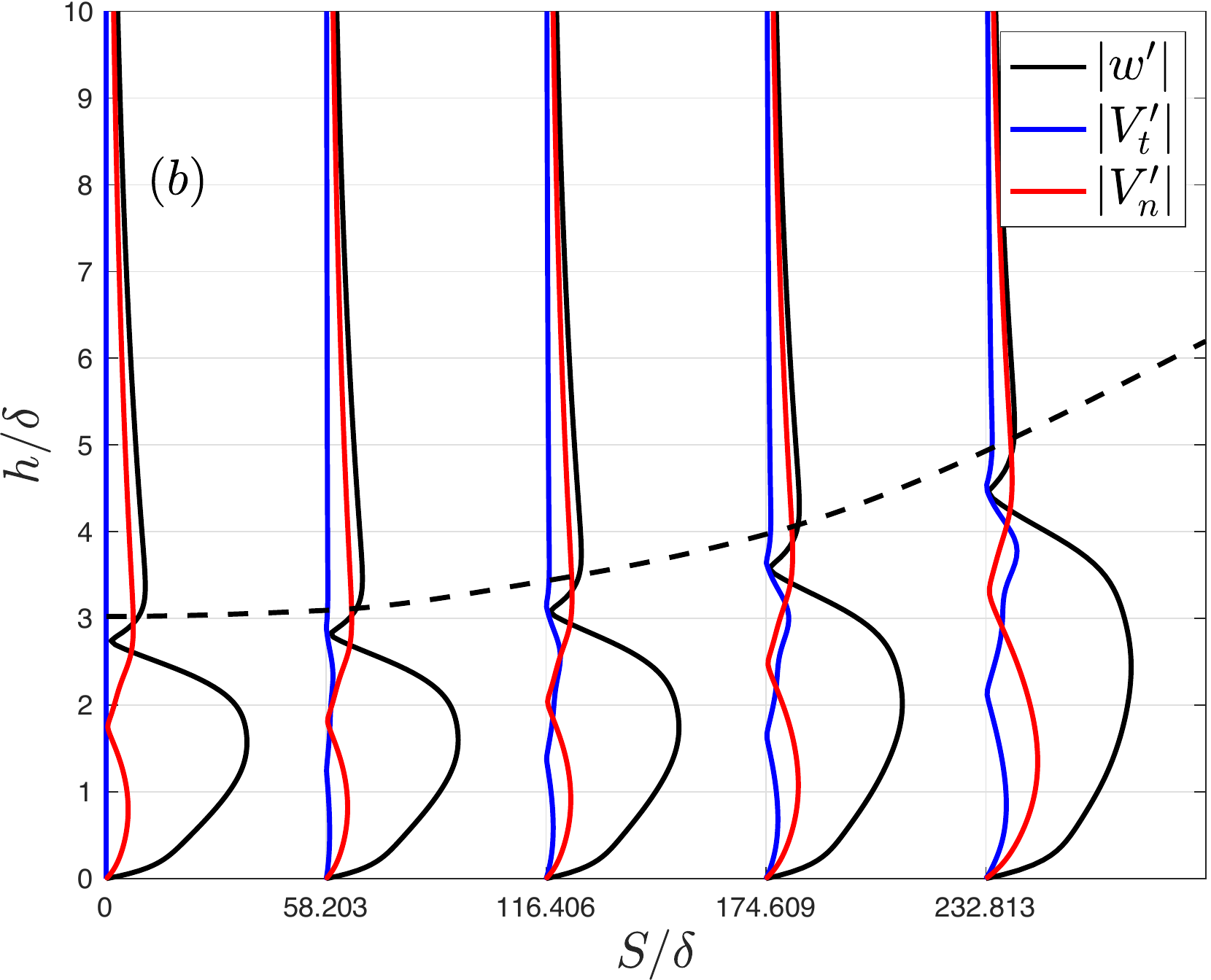}
\end{tabular}
\end{center}
\caption{Perturbation profiles along the surface \(s\) direction at (from left to right) \(s=0,58.203,116.406,174.609,232.813\) for C3376a case. \((a)\) The perturbation profiles of spanwise velocity \(|w'|\), temperature \(|T'|\) and density \(|\rho'|\). \((b)\) The perturbation profiles of spanwise velocity \(|w^{\prime}|\) surface tangential velocity \(|V^{\prime}_t|\) and wall-normal velocity \(|V^{\prime}_n|\). The dashed black lines represent the thickness of boundary layer \(\delta^{*}_{0.99}/\delta\)}
\label{Fig13}
\end{figure}

In global analysis, the temporal behavior is reflected in the eigenvalues \(\omega = \omega_r + i\omega_i\) whose imaginary part shows whether the perturbation grows or decays with time. The spatial behavior is represented by the eigenfunctions. Perturbation profiles at different \(s\)-surface locations are shown in figure \ref{Fig13}. Among all the perturbations, the temperature and density have the maximum amplitudes. The velocity perturbations, though having much smaller amplitudes, are critical for the transport of low- and high-momentum fluid. Together with the development of the boundary layer, perturbations move away from the wall. From the attachment-line region to further downstream location, both the amplitude and the affected area of velocity perturbations grow (figure \ref{Fig13}). 

\begin{figure}
\begin{center}
\includegraphics[width=0.55\textwidth]{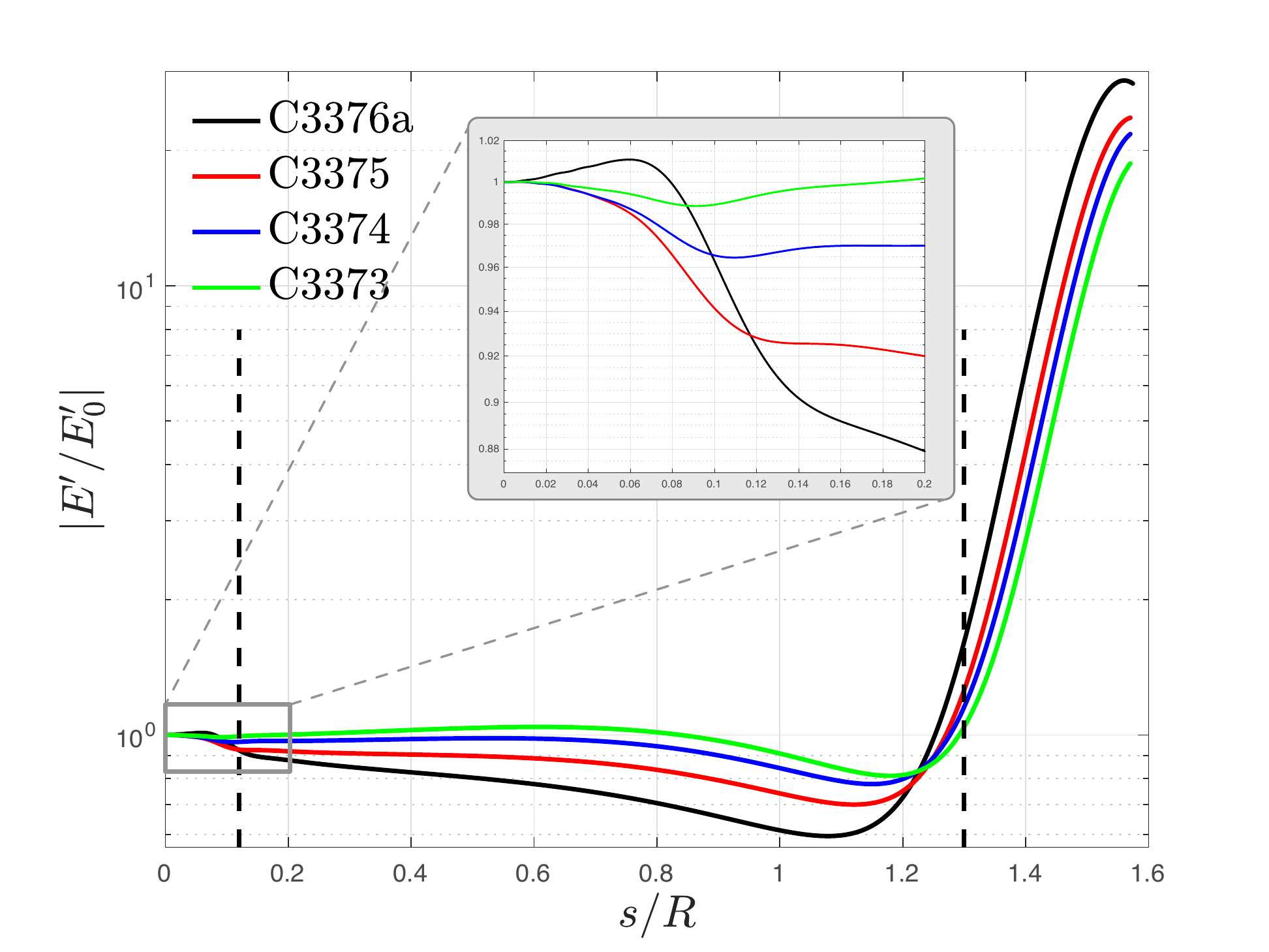}
\end{center}
\caption{Variations of the velocity perturbations norm \(|E^{\prime}|\) with respect to chord-wise location \(s/R\) for four cases. The energy norm is normalized with the energy \(|E^{\prime}_0|\) at attachment-line (\(s/R = 0\)). \(h\) represents the distance away from surface. The leading edge region is enlarged for clarity.} 
\label{Fig14_15}
\end{figure}

To further analyze the spatial behavior of perturbations, an energy norm \(E^{\prime}\) at specific \(s\) is defined for the analysis of the leading boundary layer mode. The energy norm is defined as
\begin{equation}
E^{\prime} = \int_{h} \left(\vec{\phi}^{\dag}\mathbf{M}\vec{\phi} \right)dh,
\end{equation} 
where \(\mathbf{M}\) is the energy weight matrix, the superscript \(\dag\) represents the 
conjugate transpose and \(h\) the wall normal distance. The weight matrix \(\mathbf{M}\) was originally proposed by \citet{Mack1984} and later independently derived by \citet{Hanifi1996}. It is defined as
\begin{equation}
\mathbf{M} = \text{diag}\left[\frac{T}{\gamma \rho M^2_{\infty}},\rho,\rho,\rho,\frac{\rho}{\gamma (\gamma - 1) T M^2_{\infty}} \right].
\end{equation}
According to the norm definition, both kinetic energy and the thermodynamic energy of the perturbations are taken into account. The energy norms for the four most dangerous cases are shown in figure \ref{Fig14_15}. Ignoring the influence of the outflow region, this figure shows that the development of perturbations along the surface can be divided into three regions. For the leading-edge region \(s/R\in\left[0,0.12 \right]\), as seen more clearly in the subfigure, the perturbations show, approximately, an exponential decay except for the case C3376a with $M_s=5.8$ which gives a typically algebraic growth at the region \(s/R\in[0,0.06]\). Downstream at \(s/R \in \left[0.12,1.3\right]\) is a transition region before the third region \(s/R \in \left[ 1.3, 1.57 \right]\) where the perturbations grow exponentially.

\begin{figure}
\begin{center}
\begin{tabular}{c}
{\begin{overpic}[scale=0.06,tics=5]{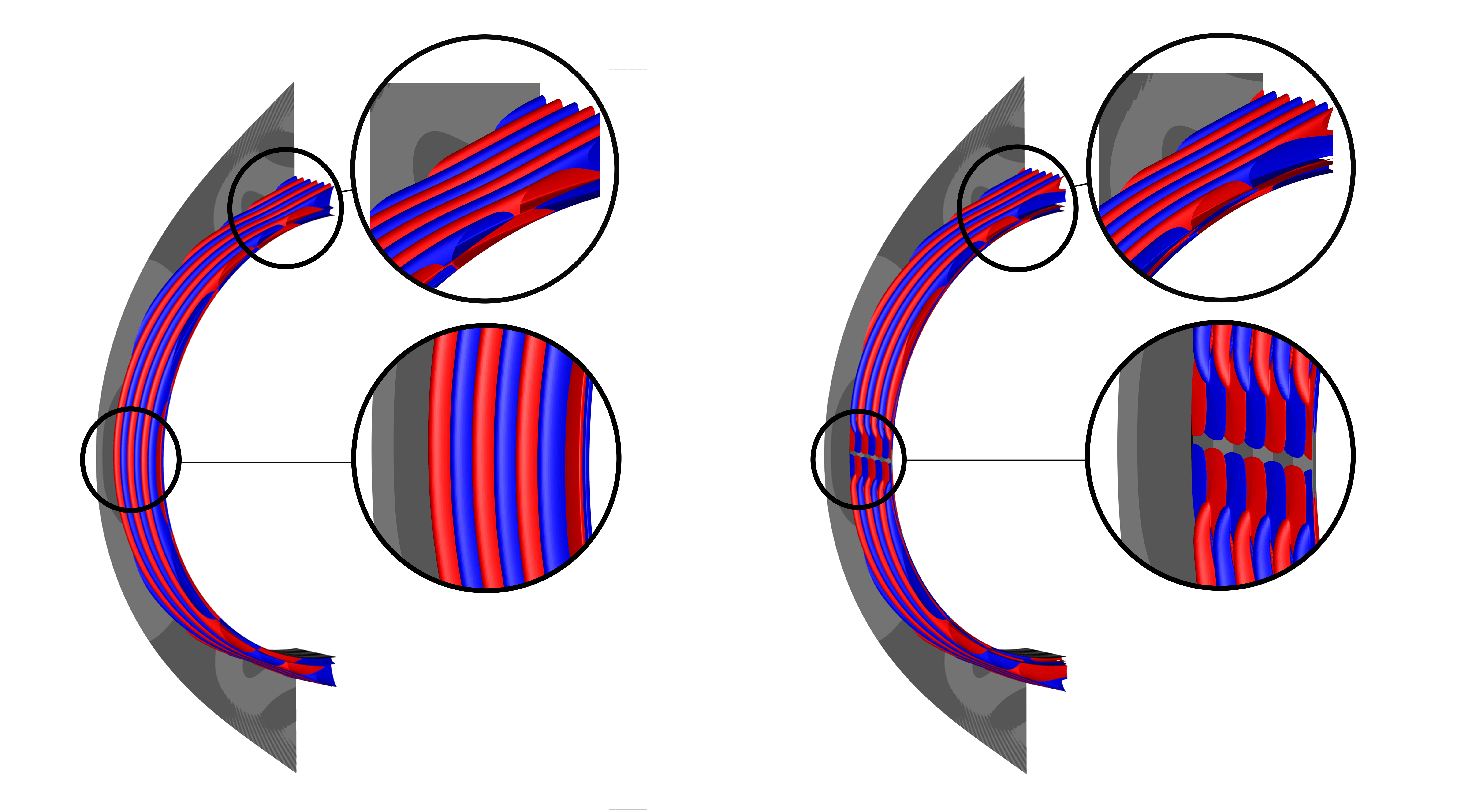}
\put(5,50){\((a)\)}
\put(55,50){\((b)\)}
\put(30,5){\(W'\)}
\put(80,5){\(V_t^{\prime}\)}
\end{overpic}} \\
{\begin{overpic}
[scale=0.262,tics=5]{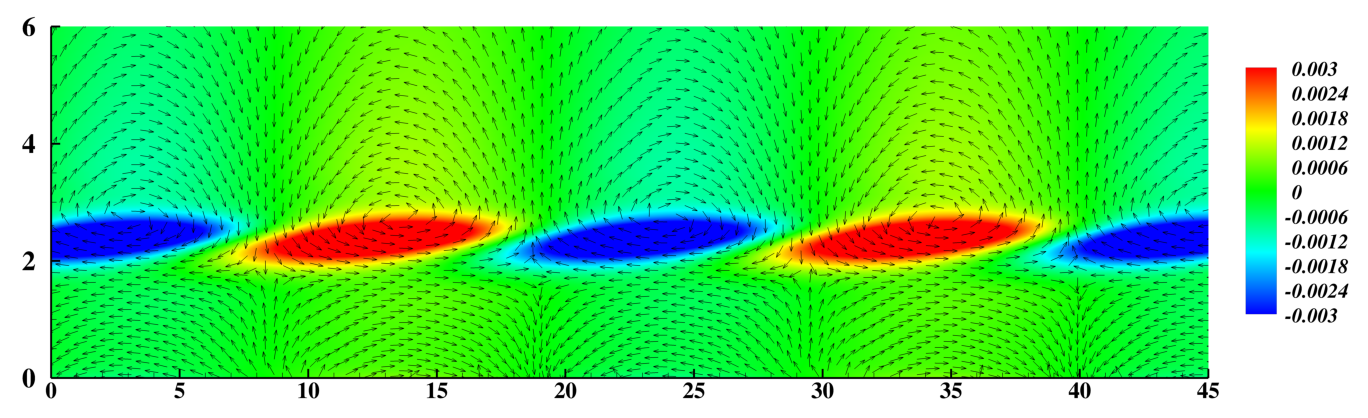}
\put(-2,25){\((c)\)}
\put(45,0.2){\(z\)}
\put(93,27){\(\rho^{\prime}\)}
\put(-2,14){\(h/\delta\)}
\end{overpic}} \\
{\begin{overpic}
[scale=0.209,tics=5]{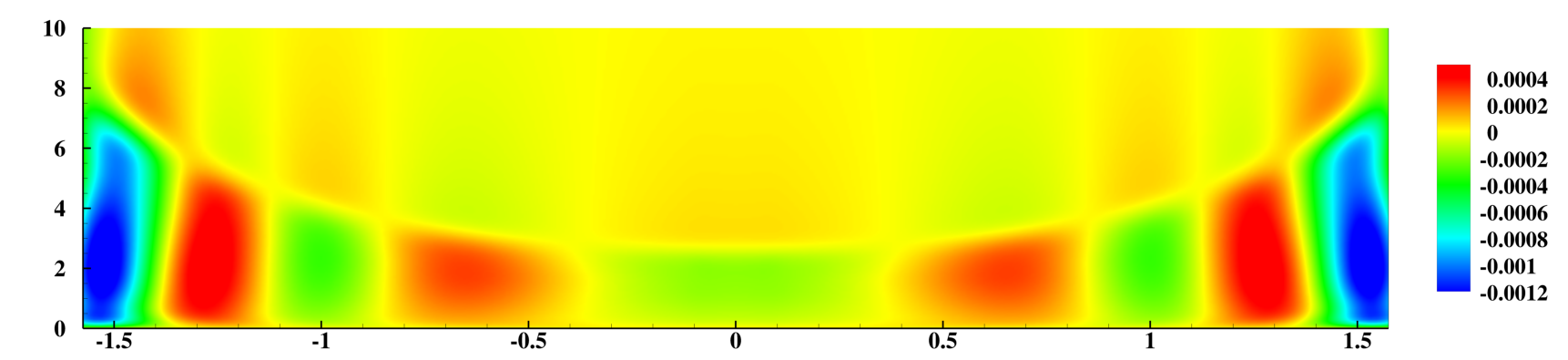}
\put(-2,20){\((d)\)}
\put(45,-1.5){\(s/R\)}
\put(93,20){\(W^{\prime}\)}
\put(-2,11){\(h/\delta\)}
\end{overpic}}
\end{tabular}
\end{center}
\caption{The leading global modes of the eigenvalue \(\omega = (0.25097, 0.00073146)\) visualized by iso-surfaces (positive value in red, negative value in blue) of \((a)\) the spanwise velocity perturbation $W^{\prime}(x,y,z) = \Re({w^{\prime}(x,y)\left(\cos \beta z + \text{i}\sin \beta z \right)})$ at contour level of \(\pm 10^{-5}\) and \((b)\) the surface tangential velocity perturbations at contour level of \(\pm 10^{-6}\), contours of the relative density perturbation are also shown at the background. \((c)\) Contour of the \(x-z\) plane cross-cut at \(y=0\) for density perturbation \(\rho^{\prime}(x,y,z)\) together with the velocity vector (unit vector) on this plane. \((d)\) Contour of the spanwise velocity perturbation $W^{\prime}$ on the \(s-n\) plane at \(z=0\).}
\label{Fig16_17}
\end{figure}
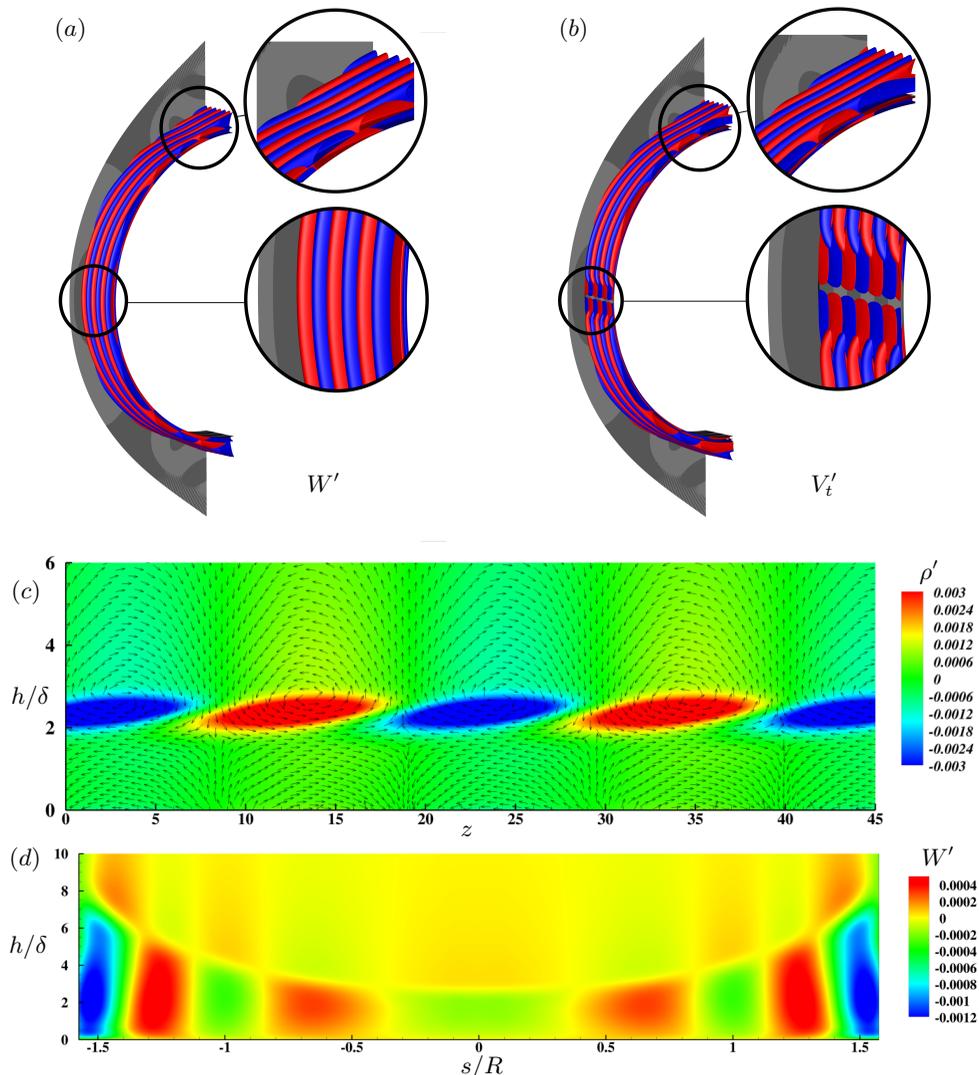



\begin{figure}
\begin{center}
\begin{tabular}{cc}
\includegraphics[width=0.49\textwidth]{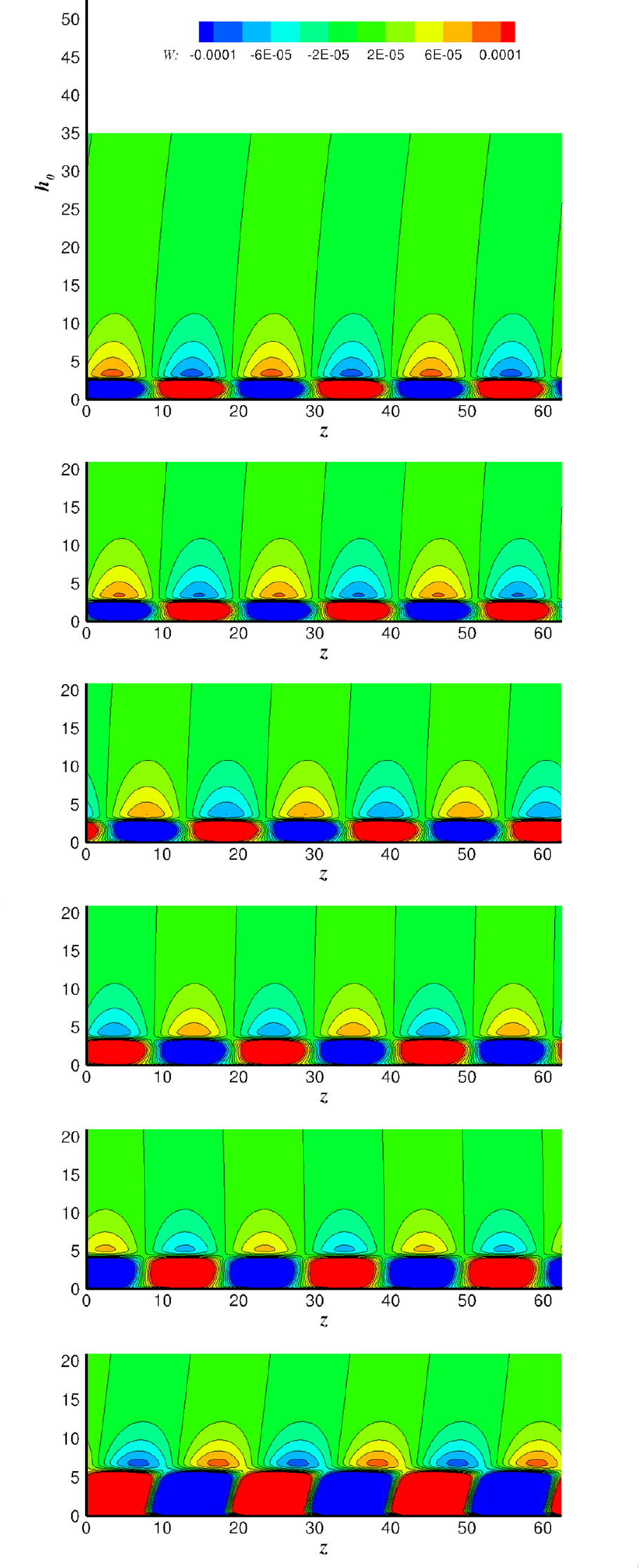} &
\includegraphics[width=0.49\textwidth]{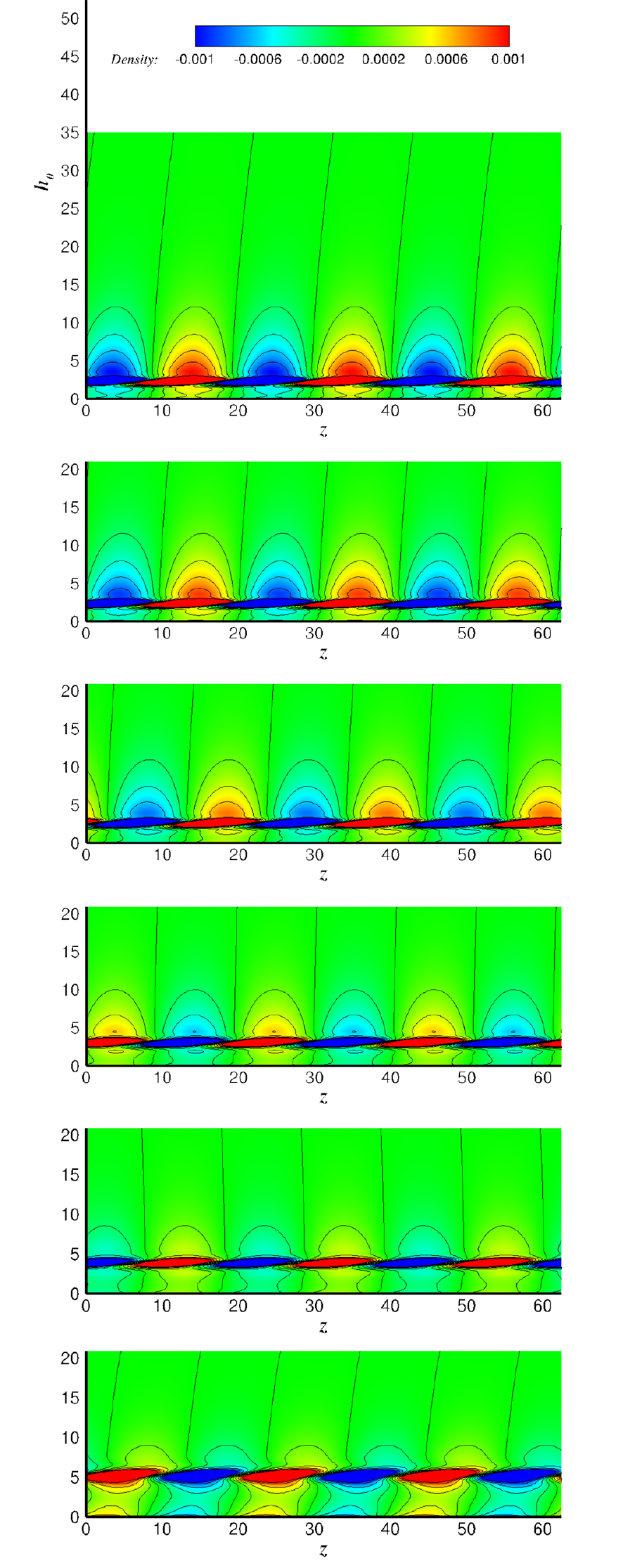}
\end{tabular}
\end{center}
\caption{Contours of spanwise perturbation \(w'\), on left hand side column, and density perturbation \(\rho'\), on the right hand side, of \(n-z\) plane along surface \(s\) direction at (from the top down) \(s=0,58.2,116.4,174.6,232.8,291.0\). \(h_0\) represents the distance away from surface.}
\label{Fig18}
\end{figure}

A 3D visualization of the perturbation $\vec{\phi}$ from the leading global eigenfunctions for C3376a case is illustrated in \(\vec{\phi}_{3D}\) as
\begin{equation}
\vec{\phi}_{3D}(x,y,z) = \Re \left[ \vec{\phi}(x,y)\left(\cos(\beta z) + i\sin(\beta z) \right)\right],
\end{equation}
where $\Re\left(\lambda\right)$ represents the real part of a complex variable $\lambda$. The three-dimensional eigenfunctions \(\vec{\phi}_{3D}\) are shown in figure \ref{Fig16_17}$(a)$ and $(b)$. 
The typical symmetric and antisymmetric structures for spanwise and chordwise velocity perturbations ($W^{\prime}$ and $V_t^{\prime}$) can be observed by iso-surfaces and contour, as shown in figure \ref{Fig16_17}\((a)\), \((b)\) and \((d)\). From the figures, one can identified that from the leading edge ($s/R = 0$ in figure \ref{Fig16_17}$(d)$) to further downstream ($|s/R| > 1.2$ in figure \ref{Fig16_17}$(d)$) the leading global mode shows a transformation of locally two-dimensional instability to locally three-dimensional instability.
At the leading edge around \(y=0\), the eigenfunction has a spatial structure similar to the local attachment-line mode of sweep Hiemenz flow as first described by \citet{Lin1996}. 
Unlike incompressible cases, the counter-rotating vortices are somewhat further away from the surface as shown in figure \ref{Fig16_17}(c). 
The vortices generate chordwise velocity streaks and similar features are identified by \citet{Mack2008} for parabolic leading-edge flow at relatively low sweep Mach numbers over an adiabatic surface. 
Further downstream, the three-dimensional instability is reflected by the obvious distortions of the iso-surface as enlarged in figure \ref{Fig16_17}$(a)$ and $(b)$. From the figure, as first shown in \citet{Mack2008},  the coexistence in the same global eigenvector of the attachment-line features at leading edge and cross-flow like features further downstream is also confirmed.
The contours of \(n-z\) planes for the spanwise velocity and density are also shown in figure \ref{Fig18} and the cross-flow like features of this mode is found becoming more prominent further downstream. However, the length of the surface in the present work is not long enough as was in the previous study \citep{Mack2008}, the leading mode here has no time or space to form full cross-flow vortices.

\section{Concluding remarks}\label{S5}
The present work attempts to explain, theoretically, the instabilities of attachment line at high sweep Mach numbers in accordance with relevant experimental conditions \citep{Gaillard1999}. The analysis is performed with high fidelity realistic base flows which are obtained with a high-order shock fitting method to fully resolve the base flow and all the geometry information. Local and global stability analyses are employed to elucidate the physics of the attachment-line instability. The theoretical results match well with the experiment. This type of attachment-line mode is found belonging to the inviscid instability, in contrast to the traditional one which belongs to the viscous instability. Thus, in low speed region, the attachment-line modes can be treated as an extension of TS modes \citep{Lin1995} while in high speed region, the attachment-line modes are closer to the Mack modes \citep{Mack1975} of the inviscid instability nature. From the global stability analysis, the leading attachment-line mode is found to be connected with cross-flow like modes further away from the attachment line. As first presented by \citet{Mack2008} over an adiabatic wall surface with relative low sweep Mach number, this connection also exists over cold wall surface with correspondingly high sweep Mach numbers.

Based on the local and global analyses at the simulation conditions, we also found that the attachment-line mode is not entirely suppressed by the leading curvature unlike in the incompressible cases. For the cases at sweep Mach number \(5.8\), the growth rate of leading global mode is found slightly larger than local calculation when the spanwise wave number \(\beta\) is above 0.2084 and the global growth rates are lesser than local calculations when the spanwise wave number is lower than 0.2084. This finding indicates that the leading edge curvature has dual effects on the attachment-line modes that it slightly destabilizes the mode for large spanwise wave numbers and stabilizes the mode for low spanwise wave numbers. 

It is also found that the more realistic base flow is the key to understand some unexplained phenomenon. The traditional boundary layer model fails to take the influence of inviscid flow into consideration, and this influence sometime may change the physics of flow instability significantly as in this case. As mentioned before, the attachment-line modes found in this study belong to inviscid mode, and the traditional attachment-line mode belongs to viscous mode. A mode competition between inviscid and viscous attachment-line modes may occur at specific parameter region, especially for lower sweep Mach number over a cold wall. Also, the inner relationship between attachment-line modes and unsteady cross-flow modes at high Mach number region is still unclear. In this study and some others \citep{Mack2008}, the attachment-line instabilities in the leading-edge region connect with cross-flow modes further away from the leading edge. Contrarily, these connections may disappear in some other cases as pointed out by \citet{Paredes2016}. The study on these aspects may be important extensions of the present research. 

\begin{acknowledgments}
Useful discussions with Dr. Zhefu Wang, Prof.\ Qibing Li of Tsinghua University and Dr.\ Jianxin Liu of Tianjin University are gratefully acknowledged. We appreciate Professors Renato Paciorri, Aldo Bonfiglioli and Xiaolin Zhong for useful discussion on shock-fitting method. Conversations with Dr.\ Pedro Paredes and Prof.\ Vassilios Theofilis on the global stability method are also helpful. This work received partial support from NSFC Grants 11602127 and 11572176, National Key Project GJXM92579, National Sci.\ \& Tech.\ Major Project (2017-II-0004-0016), NKBRPC (2014CB744801) and the Tsinghua University Initiative Scientific Research Program (2014z21020).
\end{acknowledgments}

\appendix
\section{Verification and validation of shock-fitting and boundary layer solver}\label{appA}
\begin{figure}
\centering
\includegraphics[width=0.6\textwidth]{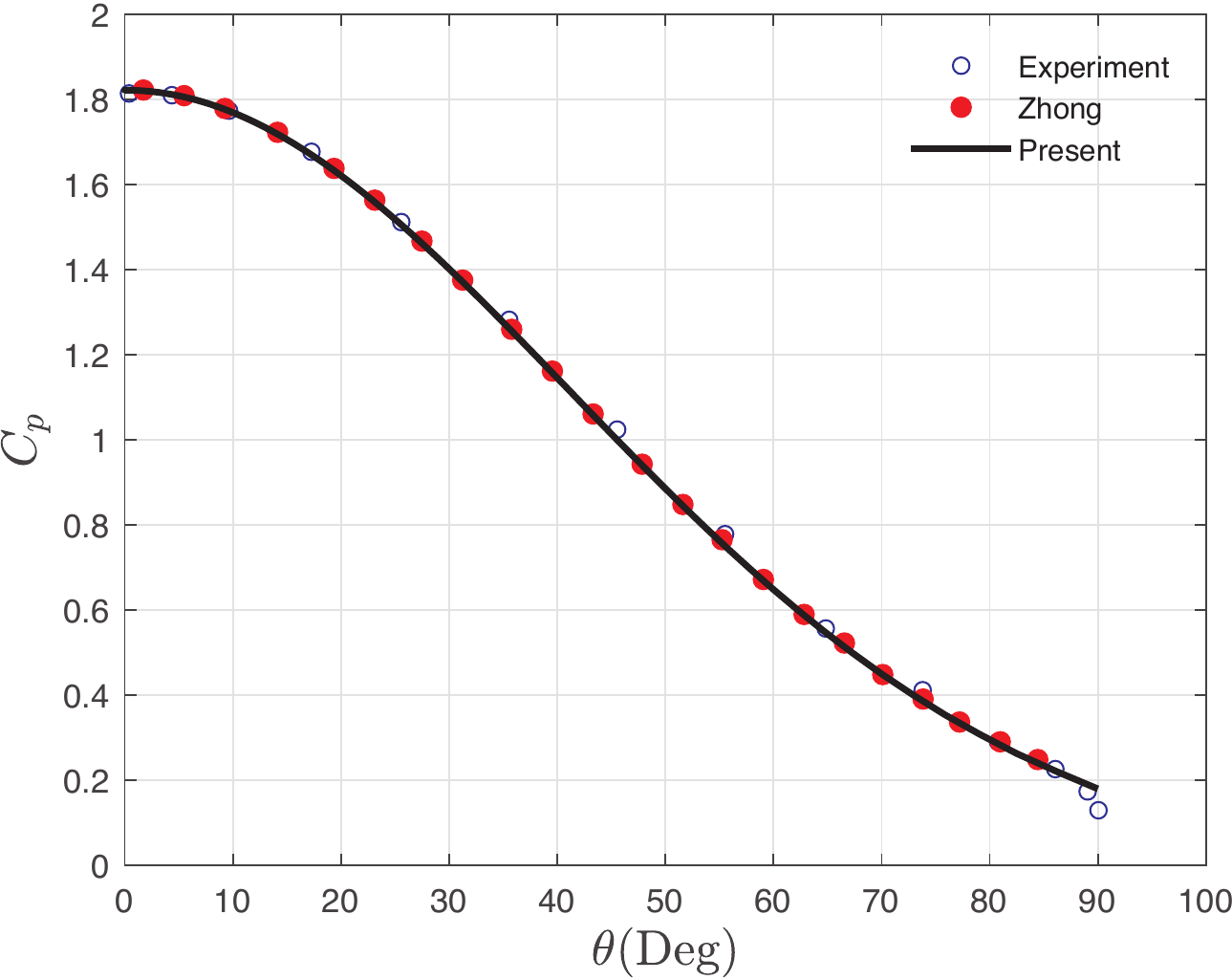}
\caption{Dependence of \(C_p\) on \(\theta\) for a flow around cylinder at Mach \(5.73\). The line represents the solution calculated with the authors' code. The circles represent results from reference and experiment \citep{Zhong1998}. }
\label{Figapp1}
\end{figure}

\begin{figure}
\centering
\includegraphics[width=0.5\textwidth]{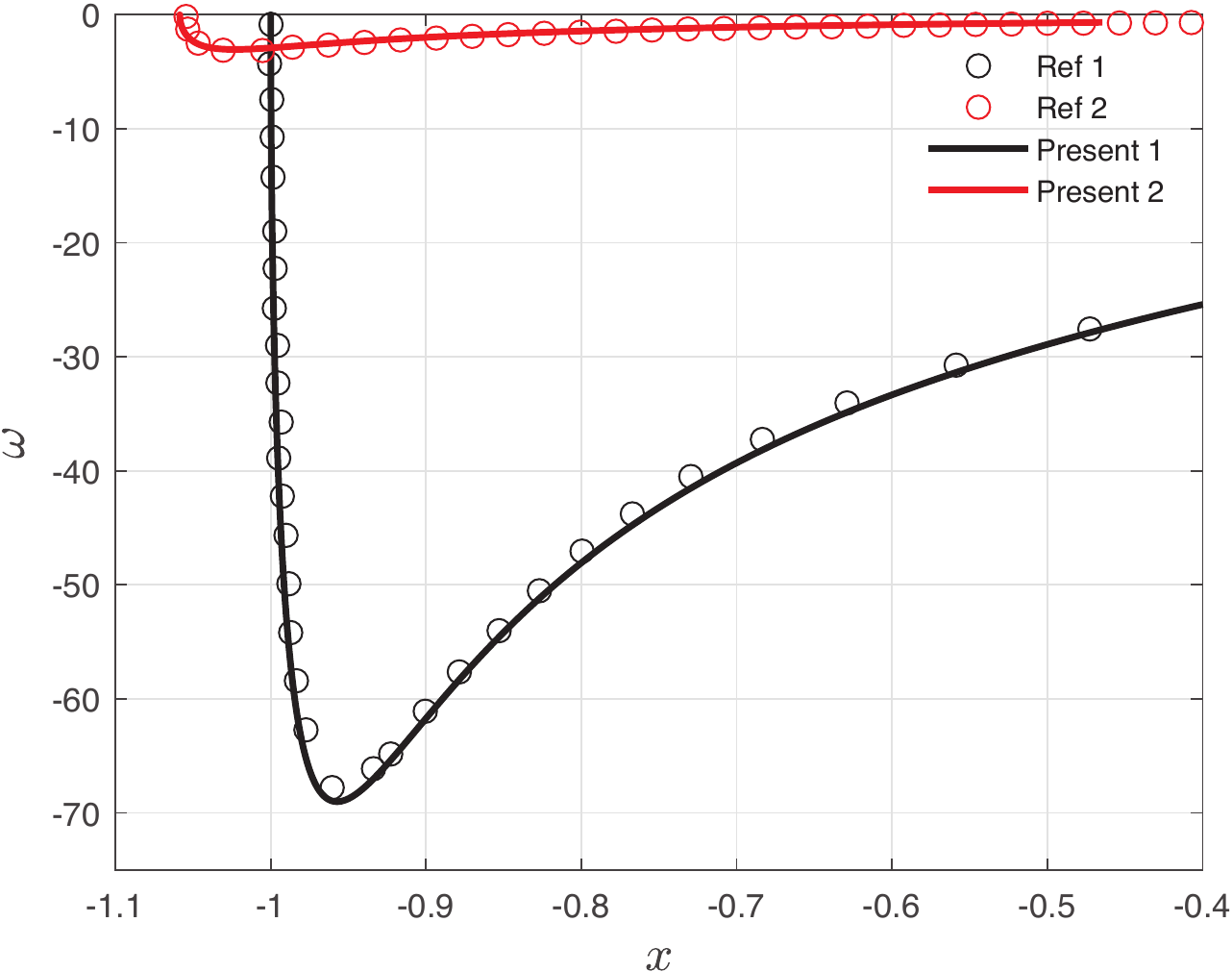}
\caption{Dependence of vorticity \(\omega\) behind the shock surface and over the wall surface over a hypersonic blunt parabola at Mach \(15\). The lines are from the solution calculated by authors' code. The circles represent results from reference \citep{Zhong1998}.}
\label{Figapp2}
\end{figure}

\begin{figure}
\centering
\includegraphics[width=0.5\textwidth]{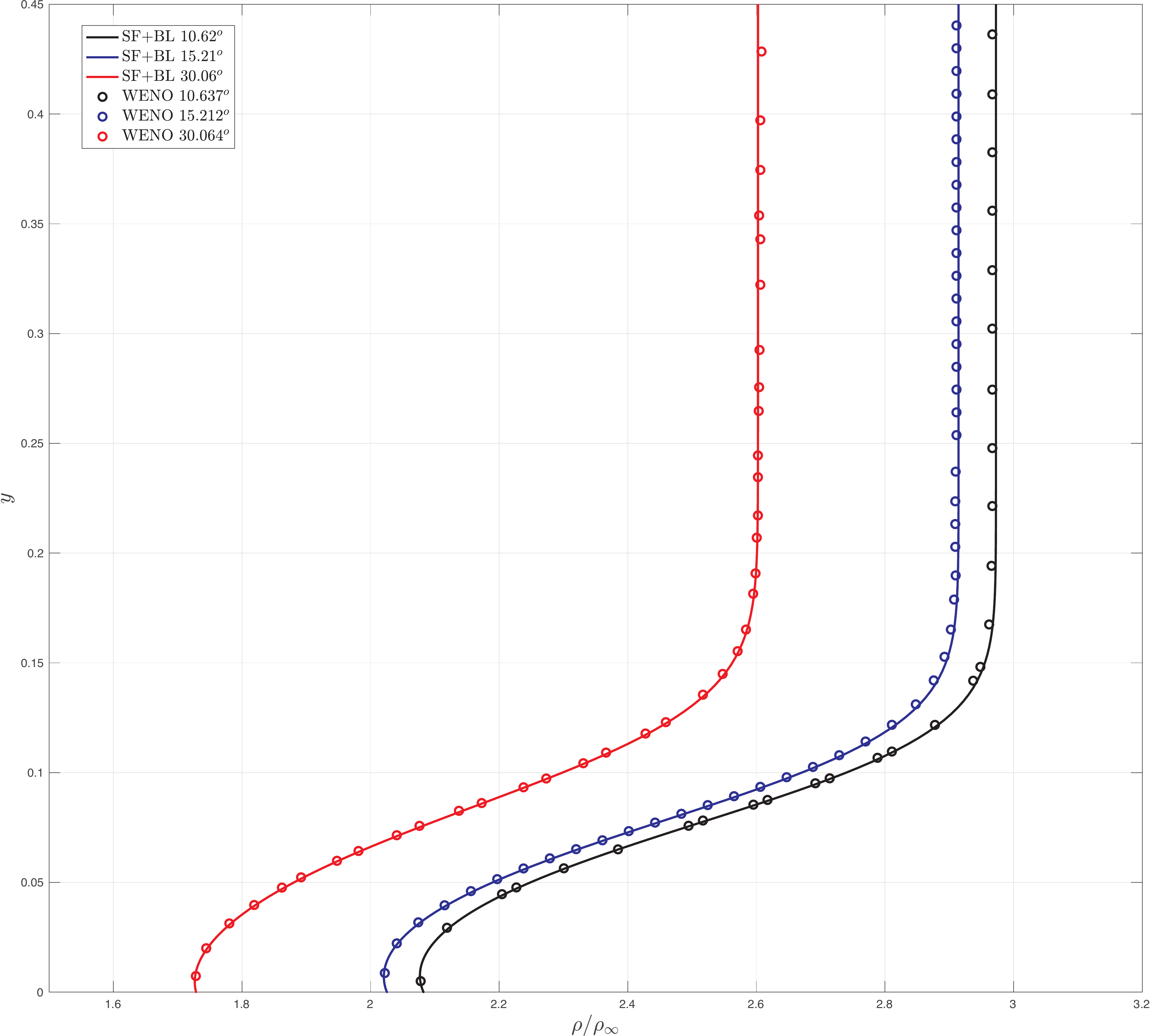}
\caption{Comparison of the density profile at several station over a sweep cylinder at Mach 3. The lines represent the solution calculated by authors' code. The circles represent the solution calculated by WENO scheme from \citet{Balakumar2012}.}
\label{Figapp3}
\end{figure}

Three cases were used to check our code. Two cases( the hypersonic flow over a cylinder and
a parabola) calculated by \citet{Zhong1998} are used for validation and verification of the present solver. Excellent agreements are achieved as shown in figure \ref{Figapp1} and figure \ref{Figapp2} for pressure coefficient and vorticities.
The last one comes from DNS study of \citet{Balakumar2012} (a supersonic flow over a sweep cylinder), we had used inviscid shock-fitting Euler solution together with boundary layer equation to solve the problem. Again, in figure \ref{Figapp3} the density profiles at several stations match perfectly.

\section{Verification and validation of global stability solver} \label{appB}
\begin{figure}
\centering
\begin{tabular}{cc}
\includegraphics[width=0.49\textwidth]{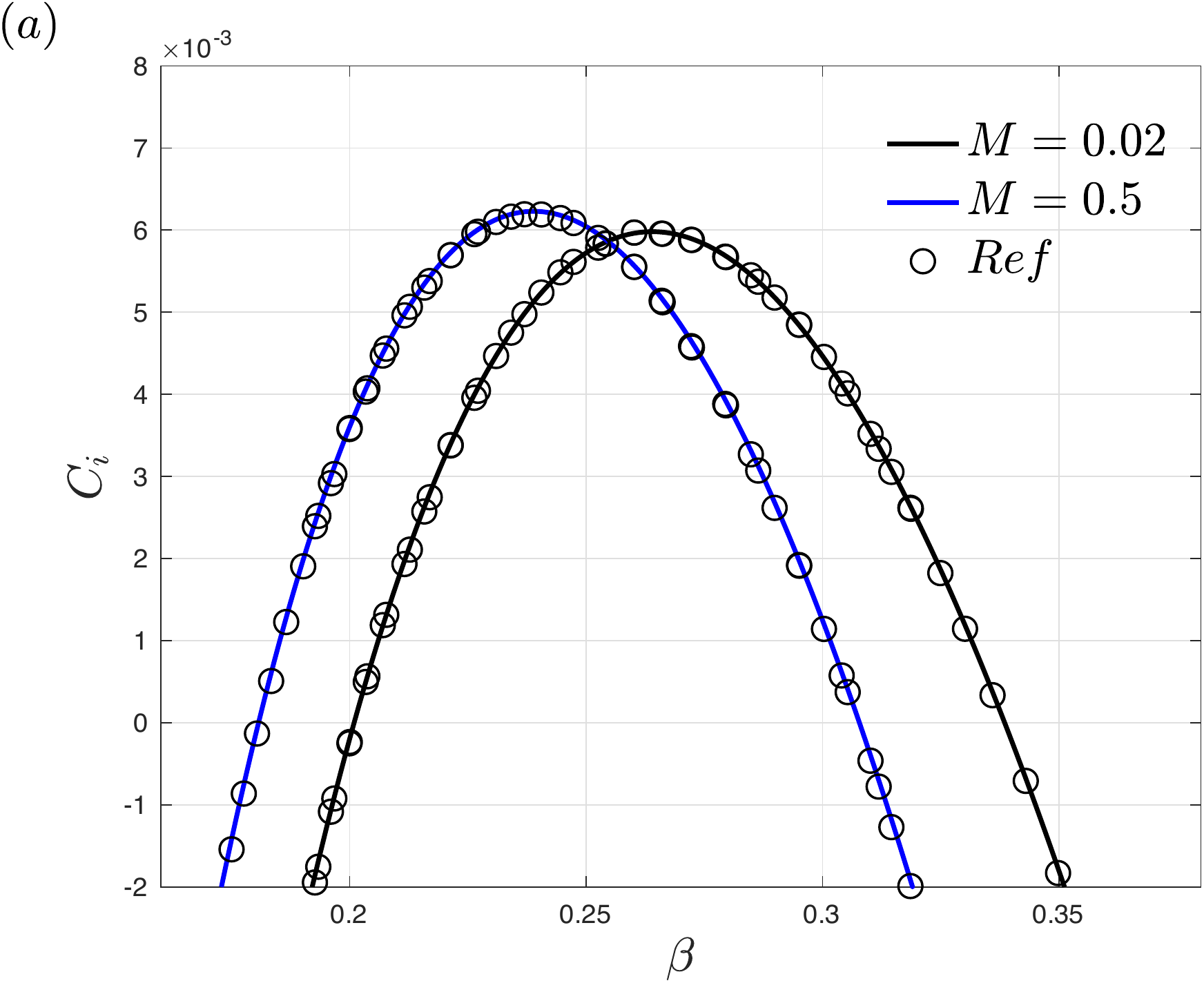}&
\includegraphics[width=0.49\textwidth]{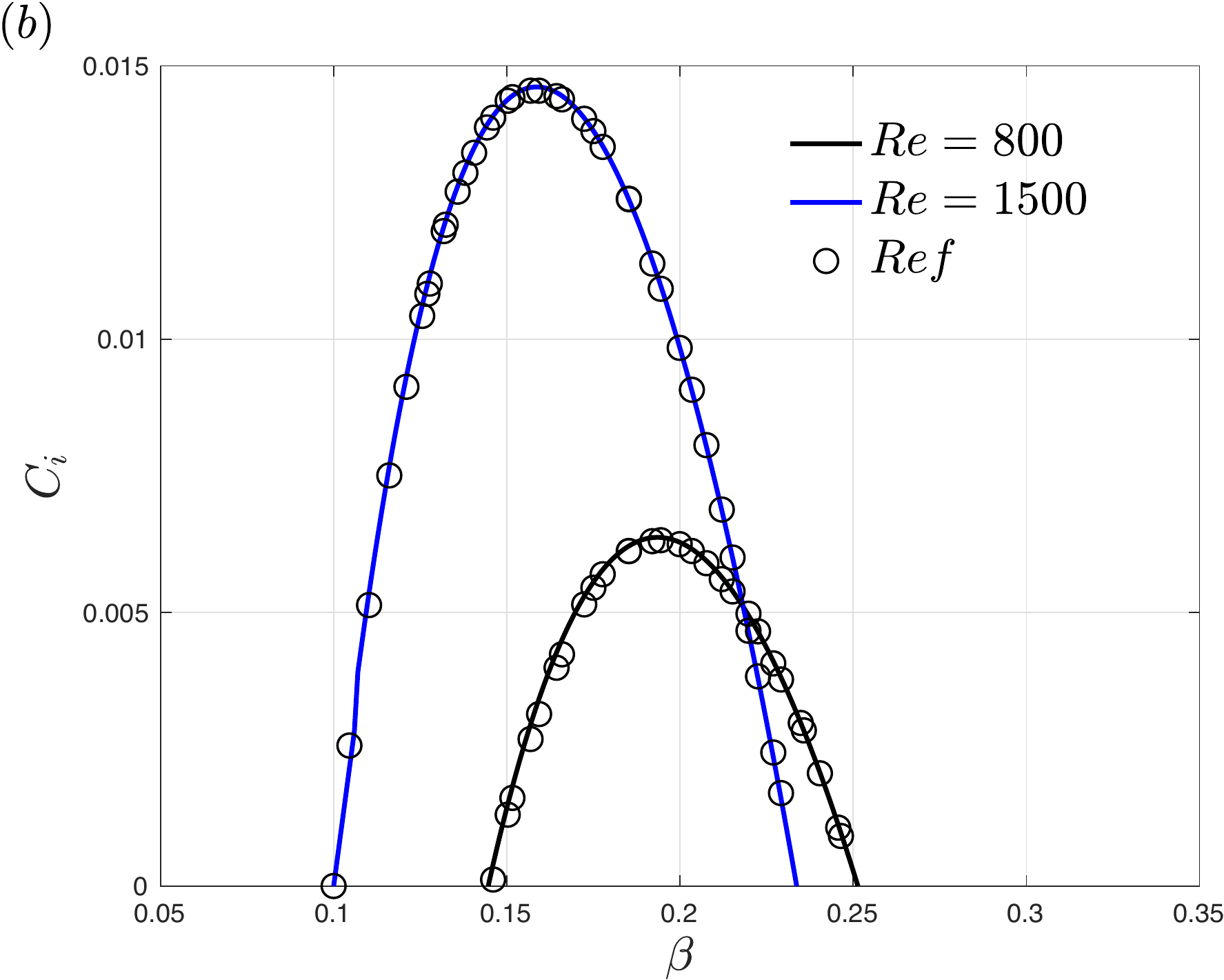}
\end{tabular}
\caption{(a).Dependence of \(C_i\) on \(\beta\) for GH mode at \(Re=800\),  
(b).Dependence of \(C_i\) on \(\beta\) for GH mode at \(M=0.9\). The data obtained by asymptotic analysis (circle) and results of the present (solid line). In this test case, we use \(121\times121\) grid points and the problem is discretized with 8th-order finite difference method.}
\label{FigAp4}
\end{figure}

Two type of cases have been used to validate the global stability solver developed in this work. First, the linear stability of the incompressible and subsonic sweep attachment line flow is addressed here to check the reliability and accuracy of the solver with the results from the literature \citep{Theofilis2006,Gennaro2013}. The dependence of the scaled eigenvalues \(C = \omega/\beta\) on \(\beta\) is shown in figure \ref{FigAp4} and these eigenvalues represents the Görlter-Hämmerlin(GH) mode of boundary layer. The boundary conditions in the present simulation keep the same as in references.

Then the solver is also compared with the local stability solver on high-speed two dimensional boundary layer cases. The spatial version of this solver is used and compared with previous study. \citet{Balakumar1992} reported an eigenvalue $\alpha=0.220-0.003091i$ for a high-speed boundary layer and the present bi-global solver gets the $\alpha=0.220199-0.003098i$. Also, for high speed boundary layer, \citet{Tumin2007} reported an eigenvalue $\alpha=0.2534420-0.0027738i$, and the present solve achieve the $\alpha=0.253442-0.002780i$. These cases are shown and compared in table \ref{TabappB}. The matches, shown in table \ref{TabappB} and figure \ref{FigAp4}, make sure the reliability and numerical accuracy of the newly developed solver.

\begin{table}
\centering
\begin{tabular}{cccc}

            & \citet{Balakumar1992} & \citet{Tumin2007} & Present Bi-Global Solver \\
Case 1  & (0.220,-0.003091)        & (0.220,-0.003091) & (0.220199,-0.003098)     \\
\hline
            & \citet{Malik1990}  & \citet{Tumin2007} & Present Bi-Global Solver \\   
Case 2  & (0.2534048,-0.0024921) & (0.2534420,-0.0027738) & (0.253443,-0.002780)
\end{tabular}
\caption{High speed boundary layer validation cases. For case 1, the parameters are as follows. The free stream Mach number $M=4.5$, the total temperature $T_0=311K$, the Prandtl number $Pr=0.72$, the Reynolds number $Re=1000$ and the frequency $\omega=0.2$. For case 2, the parameters are as follows. The free stream Mach number $M=4.5$, the total temperature $T_0=611.11K$, the Prandtl number $Pr=0.70$, the Reynolds number $Re=1500$ and the frequency $\omega=0.23$. In both cases, a \(121\times120\) grid points are used with 8th-order finite difference method}
\label{TabappB}
\end{table}

\section{Base flow solution based on boundary layer approximation}
\label{appC}
\begin{figure}
\centering
\begin{tabular}{cc}
\begin{overpic}
[scale=0.38]{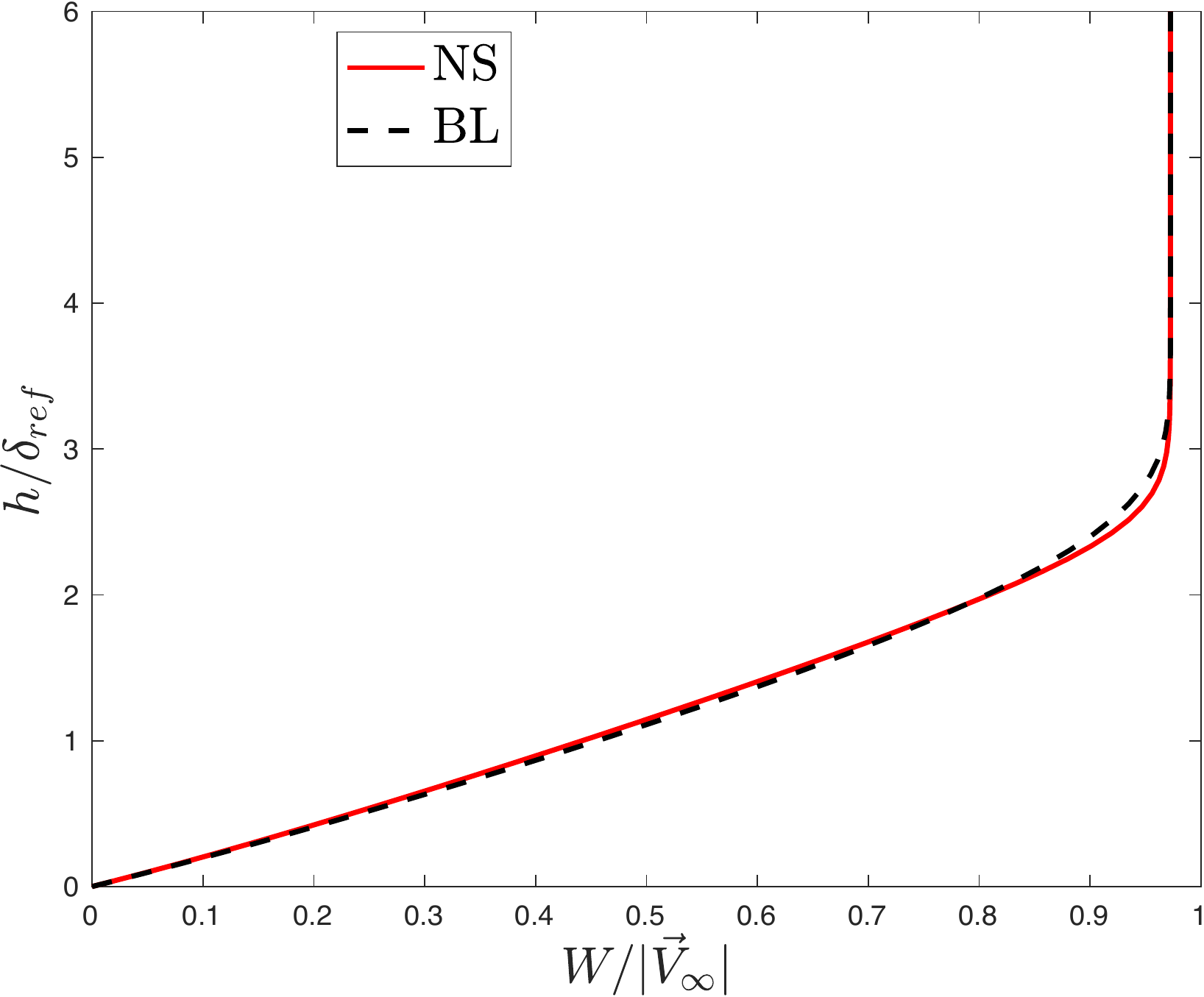}
\put(50,70){$(a)$}
\end{overpic}&
\begin{overpic}
[scale=0.38]{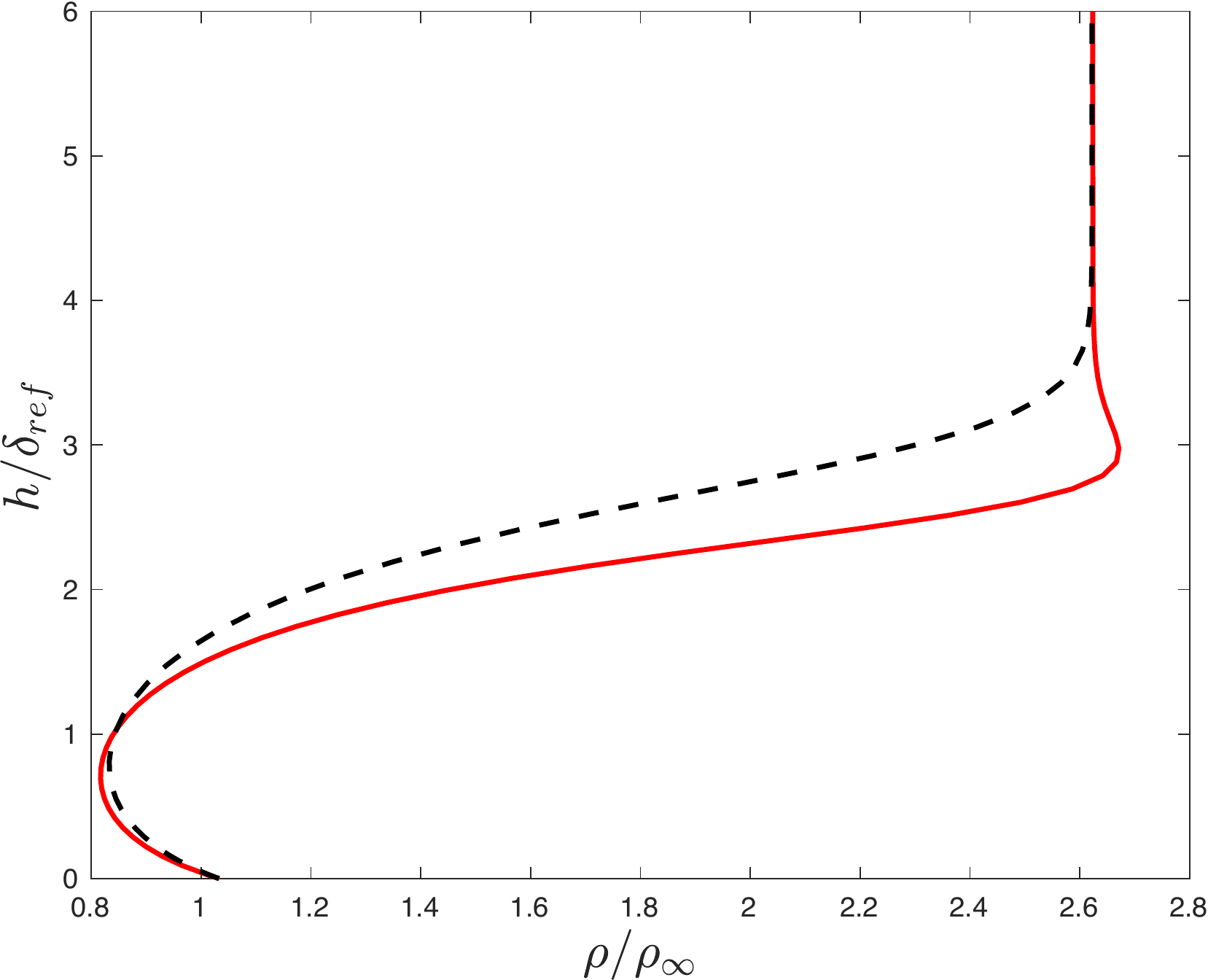}
\put(50,70){$(b)$}
\end{overpic} \\
\begin{overpic}
[scale=0.38]{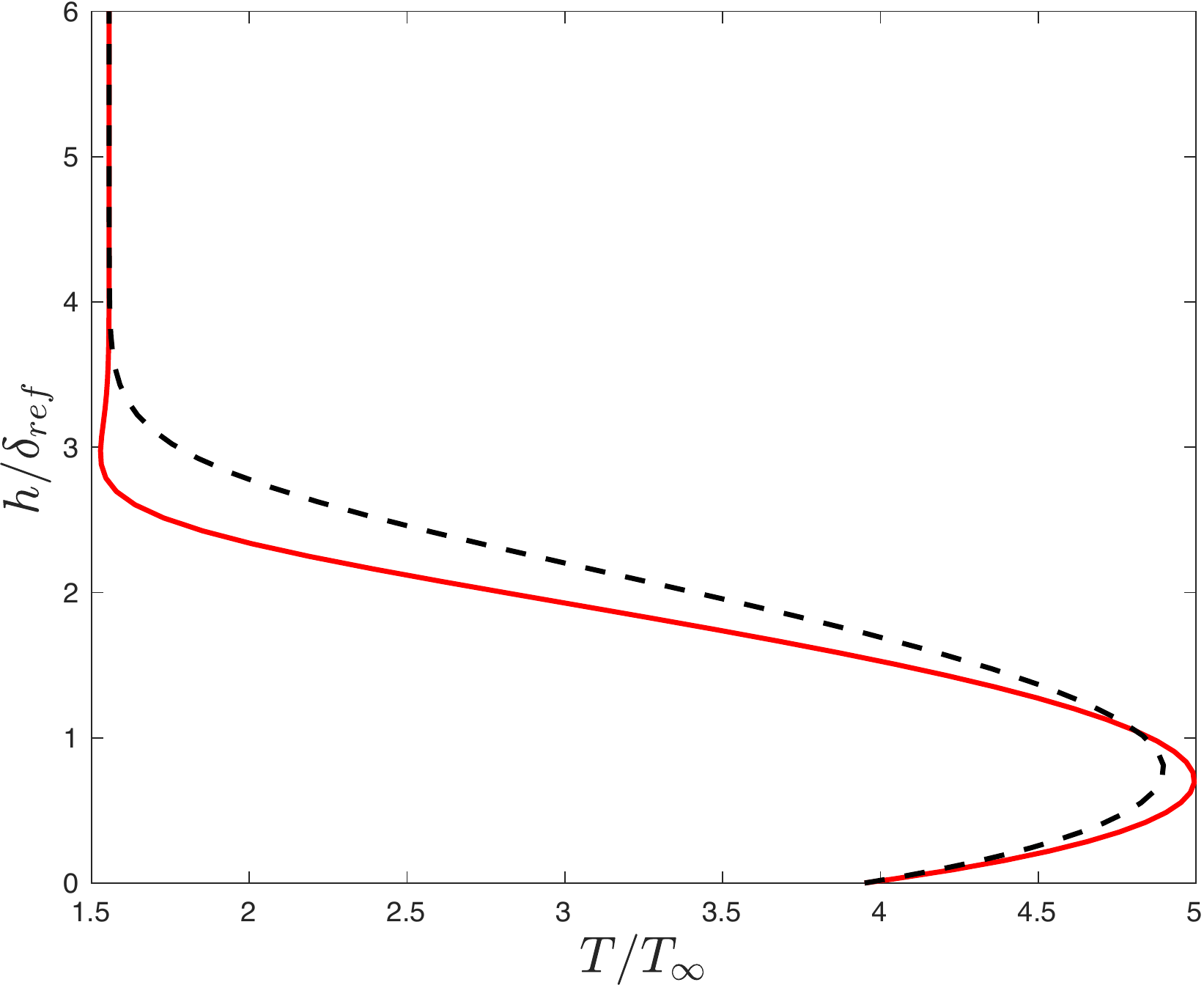}
\put(50,70){$(c)$}
\end{overpic}&
\begin{overpic}
[scale=0.38]{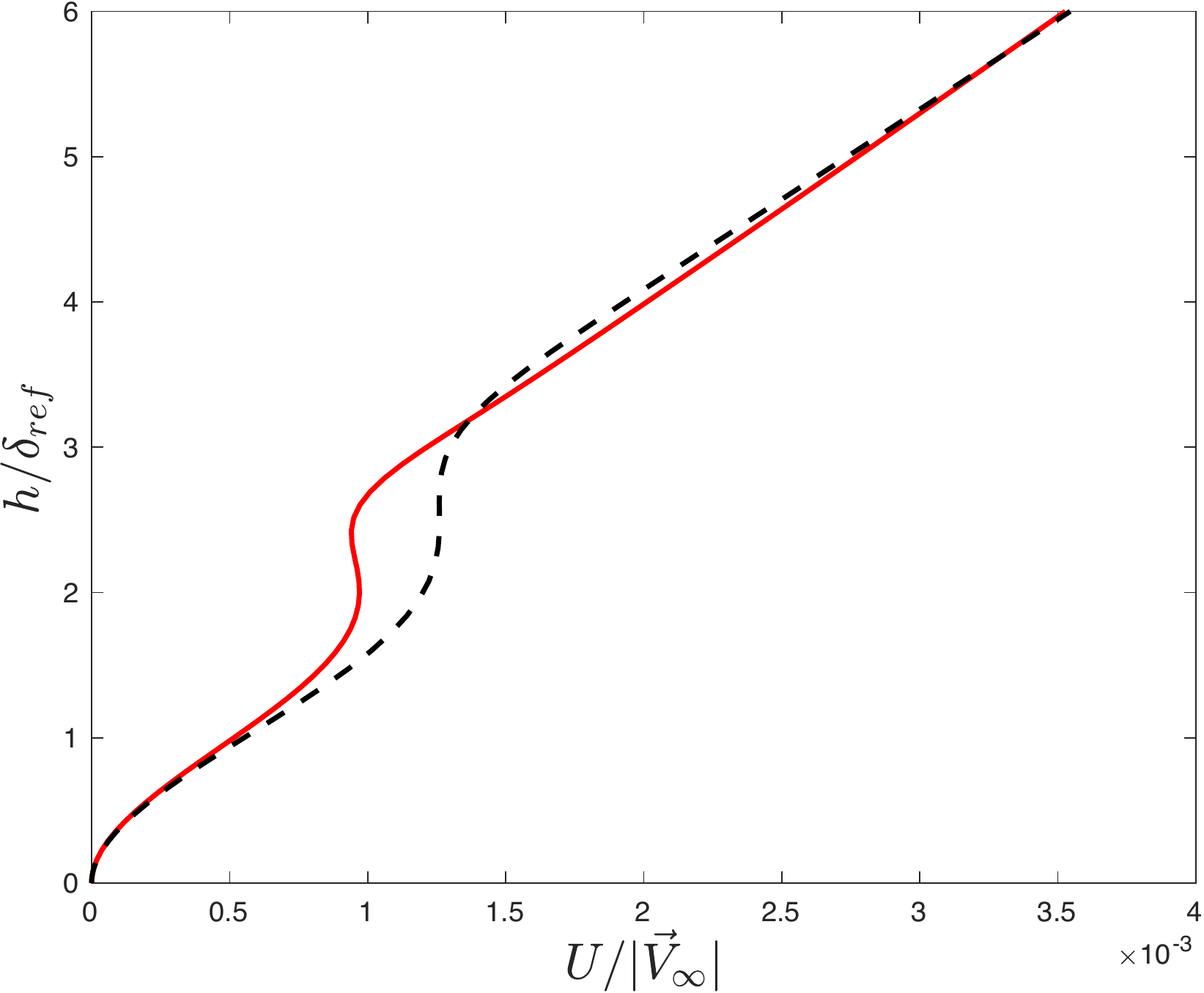}
\put(50,70){$(d)$}
\end{overpic}
\end{tabular}
\caption{The variable profiles at the attachment line. All the variables are normalized with the freestream values. Velocity is normalized with the freestream velocity \(|\vec{V}_{\infty}|\). The red line represents the solution from full NS calculation and the dashed black line is from boundary layer approximation.}
\label{FigAp5}
\end{figure}

\begin{figure}
\centering
\begin{tabular}{cc}
\begin{overpic}
[scale=0.38]{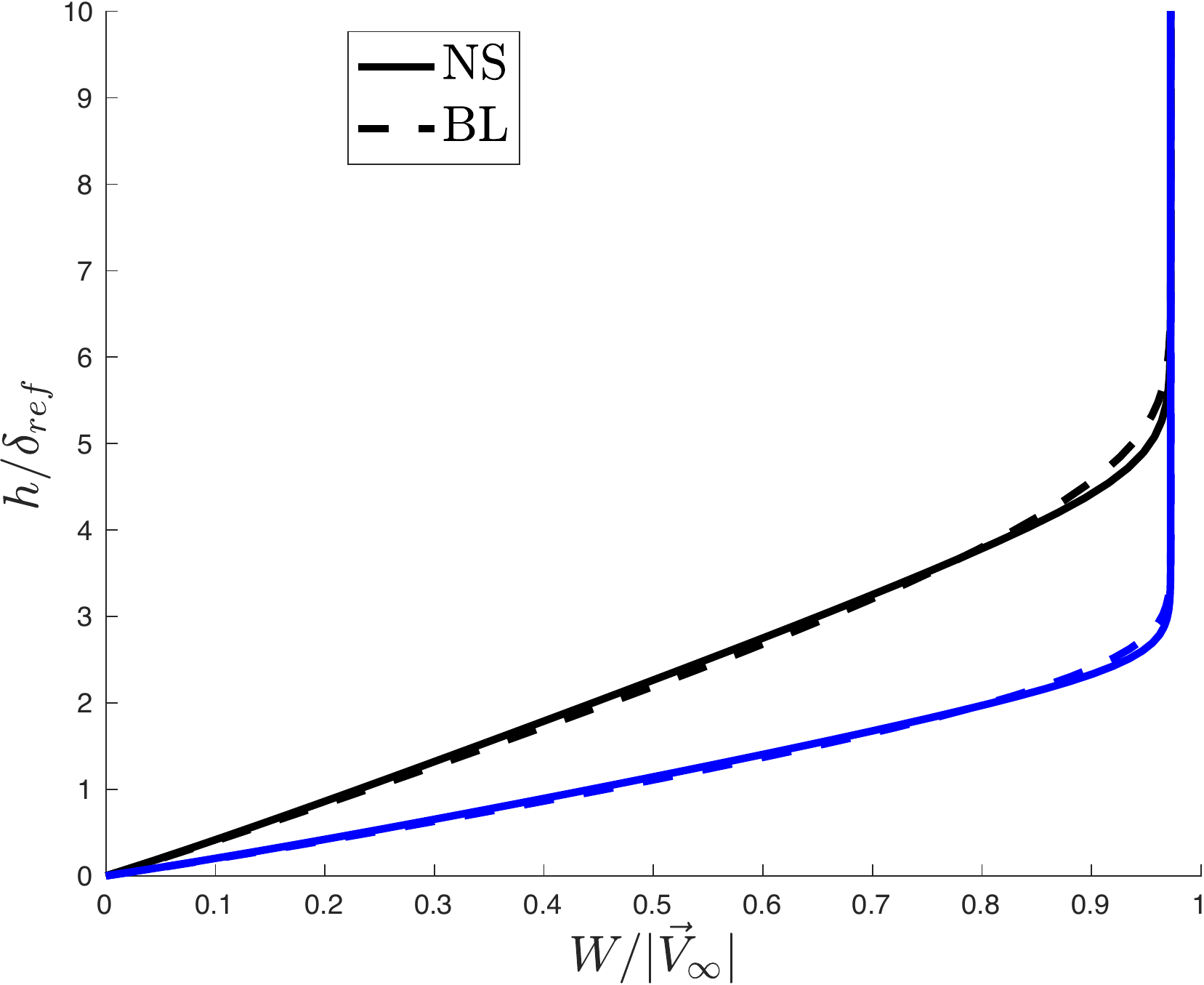}
\put(15,70){$(a)$}
\end{overpic}&
\begin{overpic}
[scale=0.38]{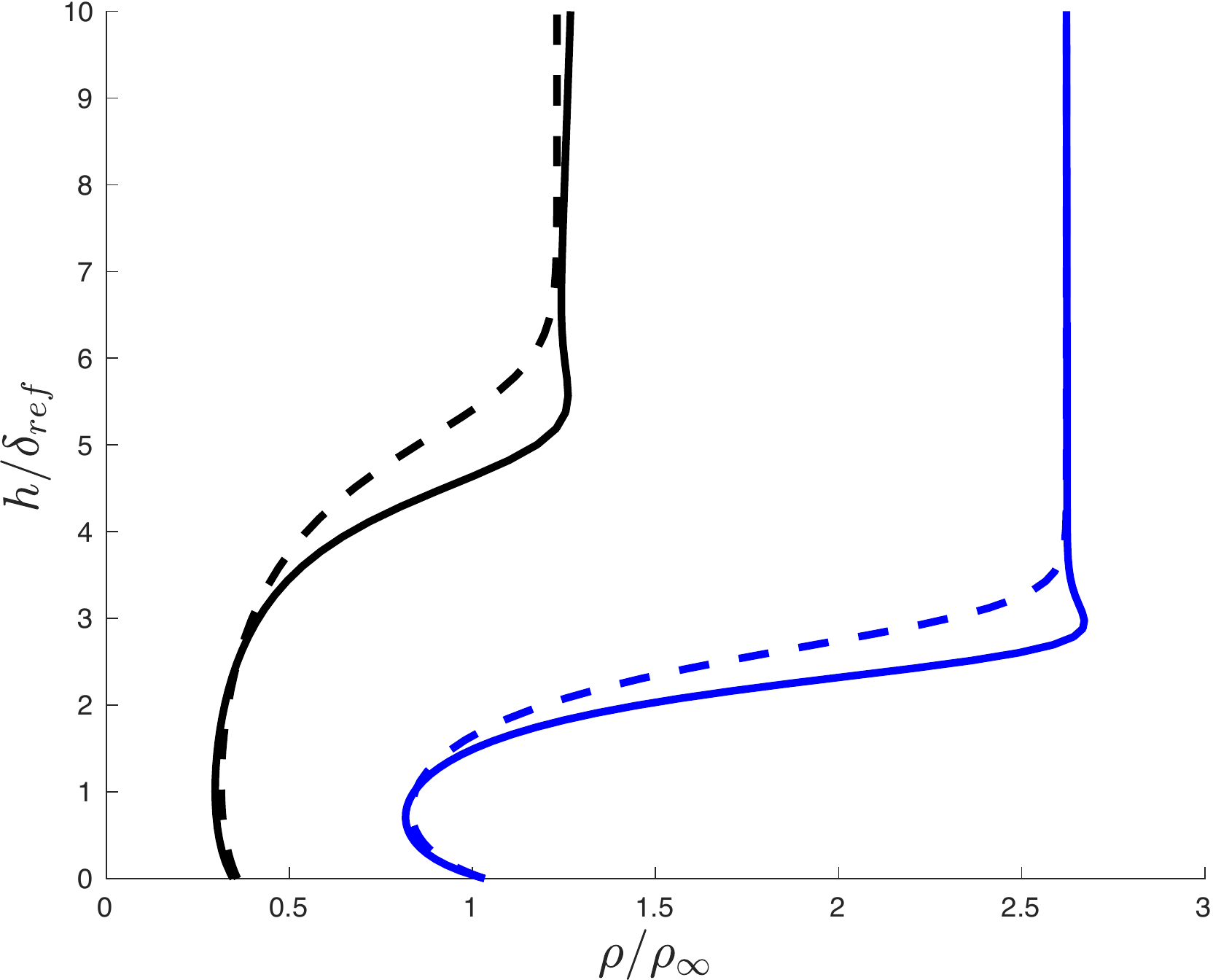}
\put(15,70){$(b)$}
\end{overpic}
\end{tabular}
\caption{The variable profiles at two different surface location: \((a)\) the spanwise velocity and \((b)\) the density. The lines represent the solution from full NS calculation and the dashed lines are from boundary layer approximation. The blue one locates at \(s = 0\) and the black one locates at \(s = 1.18R\).}
\label{FigAp6}
\end{figure}

At first, an Euler system is solved with the shock fitting method to provide the boundary information for boundary layer equations. And detailed information on boundary conditions for Euler equations can be found in \citet{Brooks2004}. Then the boundary layer equations are solved along the surface as in \citet{Wang2018}. We take the C3376a case as a typical example and other cases have similar features. At the attachment-line, the profiles for variables are shown in figure \ref{FigAp5} together with the solution from full N-S calculation. Further downstream the profiles are also shown and compared in figure \ref{FigAp6}. 

\section{\(O(1)\) equation along attachment-line}
\label{appD}
A small parameter \(\epsilon = 1/Re\) and slow variables \(y_1 = \epsilon y, t_1 = \epsilon t\) are introduced. In the framework of multiple scale approach, the perturbation is expressed as:
\begin{subequations} 
\begin{equation}
\label{MS_Expension}
\Phi(x,y,z,t) =\varphi_n\exp\left[i\beta z - i\omega t)\right],  
\end{equation}
\begin{equation}
\varphi_n=
\phi_0(x,y_1,t_1) + \epsilon \phi_1(x,y_1,t_1) + \epsilon^2 \phi_2(x,y_1,t_1) + O(\epsilon^3) + \cdots.
\end{equation}
\end{subequations}
Substituting \eqref{MS_Expension} into the linear Navier-Stokes equations, the equations for \(O(1)\) can be expressed as:
\begin{equation} \label{order_unity}
-i \omega \bm{\Gamma} \phi_0 +
 \mathbf{A}\frac{\partial \phi_0}{\partial x} +
  i\beta \mathbf{C} \phi_0 +
   \mathbf{D}\phi_0 -
    \mathbf{H}_{xx}\frac{\partial^2 \phi_0}{\partial x^2} 
- i \beta \mathbf{H}_{xz}\frac{\partial \phi_0}{\partial x}
+ \beta^2 \mathbf{H}_{zz}\phi_0 = 0,
\end{equation}
and \(O(\epsilon)\) as:
\begin{align} \label{order_epsilon}
-i \omega \bm{\Gamma} \phi_1 + \mathbf{A}\frac{\partial \phi_1}{\partial x}  + i\beta \mathbf{C} \phi_1 + \mathbf{D}\phi_1
&-\mathbf{H}_{xx}\frac{\partial^2 \phi_1}{\partial x^2} - i\beta \mathbf{H}_{xz}\frac{\partial \phi_1}{\partial x} +  \beta^2 \mathbf{H}_{zz}\phi_1
=  \\ \nonumber
&- \Gamma\frac{\partial \phi_0}{\partial t_1} + i \beta \mathbf{H}_{yz} \frac{\partial \phi_0}{\partial y_1} - \mathbf{B}\frac{\partial \phi_0}{\partial y_1} + \mathbf{H}_{xy}\frac{\partial^2 \phi_0}{\partial x \partial y_1}.
\end{align}
Looking at the equation \eqref{order_unity}, one can find that this form is the same as the form of local stability equations along a flat plate( \(z\) direction is the main stream-wise direction, \(x\) is the wall normal direction). By using the order analysis, one can find that the basic behavior along the attachment-line is govern by local theory.

\bibliographystyle{jfm}
\bibliography{Manuscript}

\end{document}